\documentclass[preprint,12pt]{elsarticle}

\usepackage{amssymb}
\usepackage{amsmath}
\usepackage{xcolor}
\usepackage{algorithm}
\usepackage{algpseudocode}
\usepackage{hyperref}
\newcommand\numberthis{\addtocounter{equation}{1}\tag{\theequation}}
\biboptions{sort&compress}
\usepackage[normalem]{ulem}

\definecolor{blue}{HTML}{000000}
\definecolor{red}{HTML}{000000}

\journal{Journal of Computational Physics}

\begin{document}

\begin{frontmatter}



\title{Implementation of integral surface tension formulations in a volume of fluid framework and their applications to Marangoni flows}


\author[inst1]{Mandeep Saini \corref{cor1}}
\cortext[cor1]{Email for correspondences: manndeepsainni@gmail.com (Mandeep Saini)}

\affiliation[inst1]{organization={Physics of Fluids Group, Max Planck Centre for Complex Fluid Dynamics, J.M. Burgers Centre for Fluid Dynamics, University of Twente},
            addressline={P.O. Box 217}, 
            city={Enschede},
            postcode={P.O. Box 217}, 
            state={State One},
            country={The Netherlands}}

\author[inst1]{Vatsal Sanjay}
\author[inst4]{Youssef Saade}
\author[inst1,inst3]{Detlef Lohse}
\author[inst2]{St\'ephane Popinet}

\affiliation[inst2]{organization={Sorbonne Université, CNRS, Institut Jean Le Rond d’Alembert},
            city={Paris},
            postcode={F-75005},
            country={France}}

\affiliation[inst4]{organization = {Canon Production Printing Netherlands B.V.},
            addressline = {5900 MA Venlo},
            country = {The Netherlands}}
\affiliation[inst3]{organization={Max Planck Institute for Dynamics and Self-Organisation},
            addressline={Am Fassberg 17}, 
            city={Göttingen},
            postcode={37077}, 
            country={Germany}}

\begin{abstract}
Accurate numerical modeling of surface tension has been a challenging aspect of multiphase flow simulations. The integral formulation for modeling surface tension forces is known to be consistent and conservative, and to be a natural choice for the simulation of flows driven by surface tension gradients along the interface. This formulation was introduced by Popinet and Zaleski \cite{popinet1999front} for a front-tracking method and was later extended to level set methods by Al-Saud \textit{et al.} \cite{abu2018conservative}. In this work, we extend the integral formulation to a volume of fluid (VOF) method for capturing the interface. In fact, we propose three different schemes distinguished by the way we calculate the geometric properties of the interface, namely curvature, tangent vector and surface fraction from VOF representation. We propose a coupled level set volume of fluid (CLSVOF) method in which we use a signed distance function coupled with VOF, a height function (HF) method in which we use the height functions calculated from VOF, and a height function to distance (HF2D) method in which we use a sign-distance function calculated from height functions. For validation, these methods are rigorously tested for several problems with constant as well as varying surface tension. It is found that from an accuracy standpoint, CLSVOF has the least numerical oscillations followed by HF2D and then HF. However, from a computational speed point of view, HF method is the fastest followed by HF2D and then CLSVOF. Therefore, the HF2D method is a good compromise between speed and accuracy for obtaining faster and correct results.

\end{abstract}

\begin{graphicalabstract}
\centering
\includegraphics{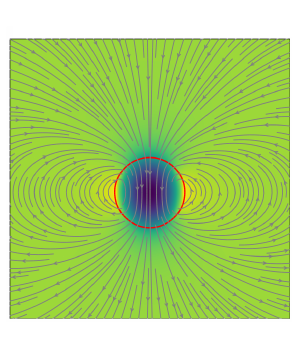}\\
\end{graphicalabstract}

\begin{highlights}
\item The integral formulation for surface tension is extended to a volume of fluid method.
\item Three different methods are proposed: a coupled level set volume of fluid method (CLSVOF), the height function method (HF), and height function to distance method (HF2D).
\item For validation, the proposed methods are tested rigorously for various problems.
\item Several new numerical benchmarks are introduced to validate the Marangoni flow solver.
\end{highlights}


\end{frontmatter}


\section{Introduction}
\label{sec:sample1}
Surface tension forces at the fluid-fluid interface originate from the cohesive interactions at the molecular level. These forces control several phenomena of multiphase flow such as breakup, coalescence, spreading, etc. \cite{eggers2008physics,eggers2024coalescence}. Surface tension can either be constant, or it can vary along the interface due to temperature gradients within the liquid, to concentration gradients within the liquid mixtures or to the inhomogeneous absorption of surfactants at the interface. In the latter case, the gradient in surface tension induces the so-called Marangoni force which leads to what is known as thermal/solutal Marangoni flow. It is associated with a plethora of interesting flow phenomenology such as convective instability responsible for bouncing droplets, Marangoni-B\'enard convection cells, tears of wine, autothermotaxis, flow instabilities in evaporating droplets, and many more \cite{scriven1960marangoni,lohse2020physicochemical,lohse2023surfactants,kant2024autothermotaxis,diddens2021competing}. Finite element methods (FEM) have been successfully used to tackle Marangoni flow problems, see for example \cite{diddens2017detailed,diddens2024bifurcation}. FEM offers many advantages, such as the implementation of monolithic solvers leading to an implicit surface tension formulation, advanced models for surfactant transport, and accurate solution construction. However, this method is non-conservative, requires the inversion of non-sparse matrices, costly regridding operation, and explicit handling of topology changes. These factors limit the applicability of such methods. A natural alternative to FEM is finite volume methods with an accurate implementation of surface tension along with an interface capturing algorithm. 
This leads us to the main contribution of the present work, where we seek such a numerical method in the framework of the volume of fluid (VOF) method.


The VOF method has several advantages over its counterparts, such as mass conservation and natural handling of topology changes. However, the numerical implementation of surface tension in VOF simulations is a difficult task. This is due to the inherent piecewise linear interface construction (PLIC), which contrasts with the continuous functional representation, necessary to precisely calculate the derivatives of interfacial shapes, and therefore to accurately compute surface tension forces. The continuous surface force (CSF) model of Brackbill \cite{brackbill1992continuum} is the most widely used method to implement surface tension force in multiphase flow simulations. However, the CSF model can suffer from inaccuracies when extended to take into account the variation in surface tension. Several implementations of the CSF model for studying Marangoni flows have been reported in the literature \cite{ma2011direct,samareh2014thermocapillary,balcazar2016level,stricker2017numerical,seric2018direct,tripathi2015dynamics}. All these works have reported errors of 4 - 10 \% in as simple a problem as Young's bubble migration \cite{young1959motion}. The main culprit for these errors is the requirement of a discrete approximation of the delta function, appearing in the surface tension term of the momentum equation, which can result in grid dependence and in slower grid convergence rates. Recently, Farsoiya \textit{et al.} \cite{farsoiya2024coupled} have proposed an improved CSF method in which they have used an uncommon definition of the discrete delta function as the ratio of the area of the interface in the computational cell to the volume of this cell. They have proposed a volume of fluid method coupled with a phase-field approach to resolve the surfactant dynamics.

As an alternative to the conventional CSF method, the integral formulation was introduced by Popinet and Zaleski \cite{popinet1999front} for implementing the surface tension in a consistent and conservative manner within the framework of a front tracking algorithm. The main advantage of this formulation is that it does not require a discrete approximation of the delta function. Recently, Al Saud et al. \cite{abu2018conservative} extended it to a level set method. Therein, the authors reported errors as low as $0.1$ \% for Young's case \cite{young1959motion}. {\color{blue} In the present work, we extend the integral formulation in a volume of fluid framework, while proposing three different algorithms and comparing them among each other.} We have implemented these methods in the open source program Basilisk, which is a partial differential equation solver on adaptive Cartesian meshes \cite{popinet2003gerris,popinet2009accurate,popinet2015quadtree}. 

This article is organized as follows: section \ref{sec:eqns} presents the governing equations, section \ref{sec:Method} describes the solution methodology. Section \ref{sec:Validation} shows the comparison among various integral formulation algorithms, along with the reference solutions (analytical solutions and numerical results from the literature) for several interfacial flow problems driven by surface tension. Finally, we draw our conclusions in section \ref{sec:conclusions} and present an outlook for future studies.

\section{Basic equations\label{sec:eqns}}
\begin{figure}
    \centering
    \includegraphics{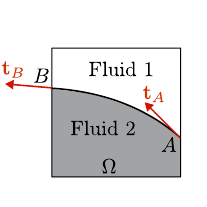}
    \caption{A representation of the control volume $\Omega$ that is cut by a fluid-fluid interface at points $A$ and $B$.}
    \label{fig:contvol}
\end{figure}
The incompressible flow of immiscible fluids separated by an interface is governed by the conservation equation for mass and momentum in each fluid,

\begin{equation}
    \partial_t \rho_i + \boldsymbol{\nabla} \cdot (\rho_i \mathbf{u}) = 0,
\end{equation}

\begin{equation}
    \centering
    \partial_t \mathbf{u}_i + \boldsymbol{\nabla} \cdot (\mathbf{u}_i \mathbf{u}_i) = \frac{1}{\rho}_i \left( - \boldsymbol{\nabla} p_i + \boldsymbol{\nabla} \cdot (2 \mu_i \mathbf{D}_i) \right),
    \label{eq:momi}
\end{equation}
along with the incompressibility constraint,
\begin{equation}
    \centering
    \boldsymbol{\nabla} \cdot \mathbf{u}_i = 0,
\end{equation}
where subscript `$i$' $\in \{1,2\}$ is used to distinguish between the phases, $\mathbf{u}$ is the velocity, $\rho$ is the density, $p$ is the pressure field, $\mu$ is the dynamic viscosity, and $\mathbf{D} = (\nabla \mathbf{u} + \nabla \mathbf{u}^T)/2$ is the deformation tensor. At the fluid-fluid interface, jump conditions are required to couple the motion between the two fluids \cite{tryggvason2011direct}. In the absence of mass transfer the velocities in both the normal and tangential directions are continuous across the interface 
\begin{equation}
    \mathbf{u}_1 = \mathbf{u}_2 \text{ at } \mathbf{x} = \mathbf{x}_I,
\end{equation}
where $\mathbf{x}$ is a position vector and $\mathbf{x}_I$ is the location of the interface. The normal stress is discontinuous across the interface to satisfy Laplace equation
\begin{equation}
    p_1 = p_2 - \gamma \kappa - 2 \mu_2 (\mathbf{n}_I \cdot \mathbf{D}_2 \cdot \mathbf{n}_I) + 2 \mu_1 (\mathbf{n}_I \cdot \mathbf{D}_1 \cdot \mathbf{n}_I) \text{ at } \mathbf{x} = \mathbf{x}_I, \label{eq:normaljump}
\end{equation}
where $\mathbf{n}_I$ is the unit vector normal to the interface, $\kappa$ is the local mean curvature of the interface and $\gamma$ is the surface tension coefficient. In case of variable surface tension, the tangential stress is also discontinuous across the interface
\begin{equation}
2\mu_1 (\mathbf{t}_I \cdot \mathbf{D}_1 \cdot \mathbf{n}_I) = \mathbf{t}_I \cdot (\boldsymbol{\nabla}_s \gamma) + 2 \mu_2 (\mathbf{t}_I \cdot \mathbf{D}_2 \cdot \mathbf{n}_I), \label{tangentjump}
\end{equation}
where $\mathbf{t}_I$ is the unit vector tangent to the interface and $\boldsymbol{\nabla}_s$ is the surface gradient. Note that the sign of the surface tension term depends on the choice of the reference fluid that is chosen here as fluid 1. 

The position of the interface is captured with a Heaviside function $(\mathcal{H}(\mathbf{x},t))$ defined as
\begin{equation}
    \mathcal{H}{(\mathbf{x},t)} = \begin{cases}
        1, \text{ if } \mathbf{x} \text{ at time } t \text{ lies in fluid 1},\ \\
        0, \text{ if } \mathbf{x} \text{ at time } t \text{ lies in fluid 2},
    \end{cases}
\end{equation}
where $\mathbf{x}$ is a position vector and $t$ is time. The interface is a material surface that evolves with the local fluid velocity $\mathbf{u}$ according to
\begin{equation}
    \partial_t \mathcal{H} + \mathbf{u} \cdot \boldsymbol{\nabla} \mathcal{H} = 0.
\end{equation} 
We use a standard one-fluid formulation to obtain a single set of conservation equations in the whole domain including the region near the interface. In this formulation, we introduce an averaged viscosity $\mu$, and an averaged density $\rho$ calculated as
\begin{equation}
    \phi = \phi_1 \mathcal{H} + \phi_2 (1 - \mathcal{H}),
\end{equation}
where $\phi \in \{\rho,\mu\}$. Thus obtained conservation equations in the one-fluid formulation are
\begin{equation}
    \centering
    \partial_t \mathbf{u} + \boldsymbol{\nabla} \cdot (\mathbf{u} \mathbf{u}) = \frac{1}{\rho} \left( - \boldsymbol{\nabla} p + \boldsymbol{\nabla} \cdot (2 \mu \mathbf{D}) + \mathbf{F}_\gamma \right),
    \label{eq:mom}
\end{equation}
\begin{equation}
    \partial_t \rho + \boldsymbol{\nabla} \cdot (\rho \mathbf{u}) = 0,
    \label{eq:density}
\end{equation}
along with the incompressibility constraint,
\begin{equation}
    \centering
    \boldsymbol{\nabla} \cdot \mathbf{u} = 0,
\end{equation}
where $\mathbf{F}_\gamma$ is the surface tension force per unit volume that appears as a source term and is non-zero only at the interface. 
Typically within the CSF formulation, surface tension is added as a singular force \cite{brackbill1992continuum},
\begin{equation}
    \int_\Omega \mathbf{F}_\gamma d\Omega = \int_\Omega (\gamma \kappa \mathbf{n}_I + \boldsymbol{\nabla}_s \gamma) \delta_s d\Omega,
    \label{eq:csf}
\end{equation}
where $\delta_s$ is the Dirac delta function which is non-zero only at the interface, and $\Omega$ is a control volume (for e.g. see figure \ref{fig:contvol}). An alternative method is proposed by Popinet and Zaleski \cite{popinet1999front}, known as the integral formulation, where the surface tension force is written as a line integral of the surface tension coefficient in the direction tangential to the interface. For an interface that intersects a control volume $\Omega$ at points $A$ and $B$, the surface tension force is simply written as
\begin{equation}
    \mathbf{F}_\gamma = \oint_A^B \gamma d\mathbf{t} = \gamma_A \mathbf{t}_A - \gamma_B \mathbf{t}_B,
\end{equation}
where $\mathbf{t}_A$ and $\mathbf{t}_B$ are the interfacial tangents at points $A$ and $B$ respectively (see figure \ref{fig:contvol}). As indicated in references \cite{popinet1999front} and \cite{popinet2018numerical}, one can show that both the CSF and the integral formulations are mathematically equivalent using Frenet's first formula for parametric curves. However, their discrete numerical implementation differs and each can have advantages as well as disadvantages, which are illustrated in table \ref{tab:diffmethods}, and are listed in what follows:
\begin{itemize}
    \item A characteristic interface thickness of the order of the mesh size $\Delta$ is required for the discrete approximation of the Dirac delta function in the CSF formulation, which is not required for the integral formulation.
    \item As a consequence of the discrete interface thickness, the order of accuracy of the CSF formulation can be sensitive to the definition of the delta function, whereas the order of convergence of the integral formulation depends only on the accuracy of the geometric description of the interface.
    \item The surface tension force in the integral formulation is locally and globally conservative, since the discrete force, for example $\gamma \mathbf{t}_A$, in figure \ref{fig:contvol}, also acts on the neighboring control volume but with a reversed sign such that the discrete integral of this force around a closed curve is zero. In contrast, the property of momentum conservation is not ensured in the CSF formulation.
    \item The tangential forces due to a spatially varying surface tension coefficient can be naturally accounted for in the integral formulation, while the CSF formulation requires the calculation of a non-trivial surface gradient term $\boldsymbol{\nabla}_s \gamma$ that can introduce inaccuracies.
    \item {\color{red} With special care, an exact balance between the pressure and surface tension terms in the momentum equation can be obtained for a static interface using the CSF formulation \cite{popinet2009accurate,popinet2018numerical}, while this property is not obvious for the discrete integral formulation. Nevertheless, Al Saud \textit{et al.} \cite{abu2018conservative} have obtained machine-accurate balance for the case of a static droplet. However, the exact mechanism of ``well balancedness" in their case remains unclear.}
    \item The extension to three-dimensional flows is not trivial for the integral formulation, but it is straightforward in the case of a CSF model. {\color{red} In three-dimensional flows, the interface is a plane, and it will intersect with the cell face along a curve. To write an integral formulations in this situation, two orthogonal tangent vectors need to be considered along the intersection curve. Additionally, it will be a challenging task to accurately calculate the intersection curve from the color function.}
\end{itemize}

\begin{table}[]
    \centering
    \begin{tabular}{|c|c|c|}
    \hline
     $~~$ & Integral formulation & CSF formulation \\ \hline
     Numerical approximation of $\delta$ function  & Not required & Required \\ \hline
     Conservative & Locally and globally & Not conservative \\ \hline
     Extension to variable surface tension & Natural & Non trivial \\ \hline
     Well balancedness & Not always achievable & Can be achieved  \\ \hline
     Extension to 3 dimensions & Non trivial & Trivial \\ \hline
    \end{tabular}
    \caption{Advantages and disadvantages of the integral formulation in comparison with the standard CSF formulation.}
    \label{tab:diffmethods}
\end{table}
\section{Numerical method \label{sec:Method}}
\subsection{Volume of fluid method \label{sec:vof}}
In the volume of fluid (VOF) framework, the Heaviside function is numerically approximated using a color function ($C$) \cite{scardovelli1999direct}. 
The color function takes a value 1 in the reference phase, 0 in the non-reference phase, and a fractional value in the cells containing the interface. There are two steps in the VOF algorithm at each time step: first the interface is geometrically constructed using piecewise linear interface construction (PLIC), then, it is advected using the local velocity field. In PLIC the line segments, described as
\begin{equation}
    \mathbf{m} \cdot \mathbf{x} = \alpha,
    \label{eq:alpha}
\end{equation}
are constructed in each cell where $0<C<1$. In equation \eqref{eq:alpha}, $\mathbf{m}$ is the local normal to the line segment, $\mathbf{x}$ is the position vector, and $\alpha$ is a unique constant for a given value of $C$ in the interfacial cell (see references \cite{scardovelli1999direct,tryggvason2011direct} for more details). After reconstruction, the interface advection equation
\begin{equation}
    \partial_t C + \boldsymbol{\nabla} \cdot (C \mathbf{u}) = C \boldsymbol{\nabla} \cdot \mathbf{u},
    \label{eq:colorfn}
\end{equation}
is solved using Weymouth and Yue's direction split advection scheme \cite{weymouth2010conservative}. Note that the color function equation \eqref{eq:colorfn} is equivalent to the density equation \eqref{eq:density}, therefore the former replaces the latter within the VOF formulation.
Equation \eqref{eq:colorfn} is discretized in each direction (shown with superscript `$d$') sequentially as
\begin{equation}
    C^{n,d+1} = C^{n,d} - \sum_f F_f(C_f,u_f,\Delta t) + \Delta t C^* (\partial_f u_f),
    \label{eq:colorfndiscrete}
\end{equation}
where subscript `$f$' denotes the cell face such that $F_f(C_f,u_f,\Delta t)$ is the flux of $C$ through face $f$ in time $\Delta t$ calculated geometrically, and $\partial_f u_f$ is the normal derivative of the face velocity. 
For more details on the advection of the color function, the reader is referred to \cite{weymouth2010conservative,arrufat2021mass}.

\subsection{Surface tension implementation \label{sec:surfacetension}}
\begin{figure}
    \centering
    \includegraphics[]{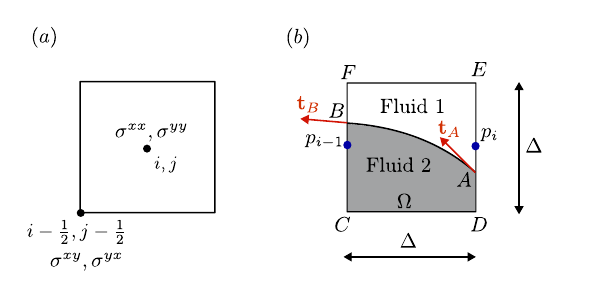}
    \caption{$(a)$ An example of a computational cell with length and width $\Delta$. In this cell, the diagonal terms of the surface tension stress tensor $\sigma^{xx}$ and $\sigma^{yy}$ are defined at the cell centers and the off-diagonal term $\sigma_{xy}$ and $\sigma_{yx}$ are defined at the vertices. $(b)$ A staggered computational cell $CDEF$ cut by the interface $AB$. Unit vectors tangent to the interface at point $A$ and $B$ are represented as $\mathbf{t}_A$ and $\mathbf{t}_B$ respectively.}
    \label{fig:defsig}
\end{figure}
In this section, we discuss the different numerical methods proposed to compute the surface tension force $\mathbf{F}_\gamma$ from the integral formulation in the VOF framework. 
We have extended the approach proposed for the level set method by Al Saud \textit{et al.} \cite{abu2018conservative}. In their approach, a discrete surface tension stress tensor $\boldsymbol{\sigma}$ is constructed such that the force in equation \eqref{eq:mom} is obtained from the divergence of said tensor,
\begin{equation}
    \int_\Omega \mathbf{F}_\gamma d \Omega = \int_\Omega \boldsymbol{\nabla} \cdot \boldsymbol{\sigma} d \Omega.
\end{equation}
The diagonal terms of $\boldsymbol{\sigma}$ are defined at the cell centers and the off diagonal terms are defined at the cell vertices (see figure \ref{fig:defsig}$(a)$). To construct the surface tension stress tensor, let us consider an example of a computational cell shown in figure \ref{fig:defsig}$(b)$ and estimate the contribution due to pressure and surface tension to the momentum equation in $x$ direction
\begin{align*}
    \frac{1}{\Omega} \int_\Omega \boldsymbol{\nabla} \cdot (- p I + \boldsymbol{\sigma}) \cdot \mathbf{e}_x d \Omega=& - \oint_{\partial \Omega} (pI \cdot \mathbf{n}_\Omega \cdot \mathbf{e}_x) ds + \oint_A^B \gamma d \mathbf{t} \cdot \mathbf{e}_x \numberthis \\
   =& - \left(\oint_D^E p ds - \oint_C^F p ds\right) + \gamma_B t^x_B - \gamma_A t^x_A, \numberthis \label{eq:pandsigint}
\end{align*}
where $\partial {\Omega}$ is the surface around the control volume $\Omega$, $\mathbf{n}_\Omega$ is the unit vector normal to $\partial{\Omega}$, $\mathbf{e}_x$ is the unit vector in the x direction, and $t^x = \mathbf{t} \cdot \mathbf{e_x}$ is the component of the tangent to the interface in $x$ direction. The pressure at the face $i \in (ED,CF)$ is discontinuous and follows Laplace's law such that

\begin{equation}
    \oint_i pds \simeq \Delta p_{i} + \Delta \gamma_i \kappa_i \begin{cases}
        s^x \text{ if } s^x < 1/2 \\
        s^x - 1 \text{ Otherwise}
    \end{cases}
    \label{eq:genpeq}
\end{equation}
where $s^x$ is the fraction of the cell face that is wetted by the non-reference phase, and $\Delta$ is the grid size. For example, in figure \ref{fig:defsig}$(b)$, $s^x = \frac{|AD|}{|ED|} < 1/2$ for the face $ED$ 
whereas $s^{x} = \frac{|BC|}{|FC|} > 1/2$ for the face $FC$. 
Upon substituting equation \eqref{eq:genpeq} in equation \eqref{eq:pandsigint} and rearranging, we obtain
\begin{equation}
    \frac{1}{\Omega} \int_\Omega \boldsymbol{\nabla} \cdot (- p I + \boldsymbol{\sigma}) \cdot \mathbf{e}_x d \Omega \simeq - \Delta (p_{i,j} - p_{i-1,j} + \sigma^{xx}_{i,j}  - \sigma^{xx}_{i-1,j}) 
\end{equation}
where the diagonal term of the surface tension stress tensor $\sigma^{xx}$ is
\begin{equation}
    \sigma^{xx}_{i,j} = \gamma_{i,j} \left[ \frac{t^x}{\Delta} + \kappa \begin{cases}
            s^x \text{ if } s^x < 1/2 \\
            s^x - 1 \text{ Otherwise}
        \end{cases}\right]_{i,j}.
        \label{eq:sigdiag}
\end{equation}
In a situation where the interface intersects with the horizontal faces, for example $CD$ or $EF$ in figure \ref{fig:defsig}$(b)$, the contribution due to the pressure jump to the $x-$momentum equation is zero, since for horizontal faces, $\mathbf{n}_\Omega \cdot \mathbf{e}_x = 0$. However, the contribution due to the surface tension term $\gamma t^x$ is nonzero. This is added to the momentum equation through the off-diagonal  terms of the stress tensor
\begin{equation}
    \sigma^{xy}_{i-\frac{1}{2},j-\frac{1}{2}} = \left[ \gamma \frac{t^x}{\Delta}\right]_{i-\frac{1}{2},j-\frac{1}{2}},
    \label{eq:signond}
\end{equation}
defined at the cell vertex $i-\frac{1}{2},j-\frac{1}{2}$. The total contribution due to pressure and surface tension to the $x-$momentum equation is
\begin{equation}
    \frac{1}{\Omega}\int_\Omega (- \nabla p + F_\gamma)  \cdot \mathbf{e}_x  d\Omega \simeq - \Delta \left( p_{i,j} + \sigma^{xx}_{i,j} - p_{i-1,j} - \sigma^{xx}_{i-1,j} + \sigma^{xy}_{i-\frac{1}{2},j+\frac{1}{2}} - \sigma^{xy}_{i-\frac{1}{2},j-{\frac{1}{2}}} \right).
    \label{eq:ffromsig}
\end{equation}
The contribution to the $y-$momentum equation can be obtained by the permutation of indices $i,j$ and superscripts in equation \eqref{eq:ffromsig}. For a detailed derivation of the surface tension stress tensor the reader is referred to the article by Al Saud \textit{et al.} \cite{abu2018conservative}.
The key elements of the stress tensor $\boldsymbol{\sigma}$ are the tangent vector $\mathbf{t}$, the curvature $\kappa$, and the face fraction $s^x$. To obtain these elements, Al Saud \textit{et al.} \cite{abu2018conservative} used the sign-distance function $d$ and proposed algorithms \ref{alg:sigxx} and \ref{alg:sigxy} which we are also going to use in our formulations. {\color{blue} Algorithm \ref{alg:sigxx} is used to compute the diagonal terms $\mathbf{\sigma}_{xx}$ based on equation \ref{eq:sigdiag}. First, the tangent and surface tension coefficient are weighted with $\xi$, which is and approximation of the surface fraction obtained from sign distance function. Then these weighted values are then used as inputs in equation \ref{eq:sigdiag}. In a similar way in the algorithm \ref{alg:sigxy}, non-diagonal terms are calculated using equation \ref{eq:signond}.} The fundamental distinction between the numerical schemes proposed in this manuscript lies in the way these key elements are calculated from the interface shape represented with a color function $C$ within the VOF framework.  We propose three different methods which are discussed next.

\begin{algorithm} [ht!]
    \caption{Calculate diagonal term $\sigma^{xx}$}
    \label{alg:sigxx}
    \begin{algorithmic}
    \Require $d,\gamma,t^x,\kappa$
    \State $\sigma^{xx} \gets 0$
        \For {$k$ in $-1,1$}
            \If{$d_{i,j} (d_{i,j} + d_{i,j+k})  {\color{red} \leq 0}$}
            \State $\xi \gets \frac{d_{i,j}}{(d_{i,j}-d_{i,j+k})}$
            \State $t^x \gets 2\xi t^x_{i,j+\frac{k}{2}}+ (1 - 2\xi) t^x_{i,j}$
            \State $\gamma \gets \gamma_{i,j} + \xi (\gamma_{i,j+k} - \gamma_{i,j})$
            \State $\sigma^{xx}_{i,j} \gets \sigma_{i,j}^{xx} + \gamma \left[\frac{|t^x|}{\Delta} - {\rm sign}(d_{i,j}) \kappa \left(\frac{1}{2} - \xi \right)\right]$
        \EndIf 
    \EndFor
\end{algorithmic}
\end{algorithm}

\begin{algorithm} [ht!]
    \caption{Calculate $\sigma^{xy}$}
    \label{alg:sigxy}
    \begin{algorithmic}
        \Require $d,\gamma,\mathbf{t}$
        \If{$(d_{i-1,j-\frac{1}{2}}) (d_{i,j-\frac{1}{2}}) > 0$}
            \State $\xi \gets \frac{d_{i-1,j-\frac{1}{2}}}{d_{i-1,j - \frac{1}{2}} - d_{i,j-\frac{1}{2}}}$
            \State $t^x \gets \xi t^x_{i,j-\frac{1}{2}}+ (1 - \xi) t^x_{i-1,j-\frac{1}{2}}$
            \State $\gamma \gets \gamma_{i-1,j-\frac{1}{2}} + \xi (\gamma_{i,j-\frac{1}{2}} - \gamma_{i-1,j-\frac{1}{2}})$
            \State $\sigma^{xy}_{i-\frac{1}{2},j-\frac{1}{2}} \gets - \gamma {\rm sign}(d_{i,j-\frac{1}{2}}) \frac{t^x}{\Delta}$
        \EndIf
    \end{algorithmic}
\end{algorithm}

\subsubsection{Coupled level set volume of fluid method (CLSVOF)\label{sec:clsvof}}
The level set coupled with VOF was proposed to overcome the shortcomings of each individual use of these two methods \cite{sussman2000coupled}. In the case of level set, the main disadvantage is the poor mass conservation, whereas for VOF, the difficulties lie in the calculation of higher order derivatives of the interface (e.g. curvature) using the color function $C$. While there are certainly benefits from the coupling of both level set and VOF, there are some additional concerns to this approach, namely, the computational cost that stems out of two simultaneous descriptions of the same interface and the intricate coupling between both.
Keeping this point in mind, we propose the following algorithm \ref{alg:clsvof} for the implementation of the integral formulation along with a CLSVOF representation of the interface.

\begin{algorithm}
\caption{Algorithm for CLSVOF method}\label{alg:clsvof}
\begin{algorithmic}
\Require $\phi^n$,$C^n$
\State Obtain $C^{n+1}$ from equation \eqref{eq:colorfndiscrete} using Weymouth and Yue \cite{weymouth2010conservative} algorithm
\State $\phi^* \gets \phi^n + \Delta t \nabla \cdot (\mathbf{u}^n \cdot \phi^n)$
\State Reconstruct VOF using equation \eqref{eq:alpha} and calculate $\phi^{VOF} = \alpha \Delta$.
\State $\phi^{**} \gets W \phi^{VOF} + (1 - W) \phi^{*}$ \Comment{$W$ is set to 0.1}
\State Calculate $\phi^{n+1}$ using redistancing  algorithm from reference \cite{limare2023hybrid}
\State $\kappa \gets \kappa(\phi^{n+1}), t^x \gets t^x (\phi^{n+1})$
\For{All cells}
\State Calculate $\sigma^{xx} (d = \phi^{n+1})$ and $\sigma^{yy} (d = \phi^{n+1})$ using Algorithm \ref{alg:sigxx}
\EndFor
\For{All vertices}
\State Calculate $\sigma^{xy} (d = \phi^{n+1})$ and $\sigma^{yx} (d = \phi^{n+1})$ using Algorithm \ref{alg:sigxy}
\EndFor
\State Calculate $F_\gamma$ using equation \eqref{eq:ffromsig} and add to the momentum equation.
\end{algorithmic}
\end{algorithm}

The shape of the interface is known at the beginning of a time step; therefore, both the sign distance (level set) and the VOF color functions $(\phi^n,C^n)$ are accurately calculated. 
The color function $C^n$ is then geometrically advected using a consistent and conservative directionally split scheme described in Section \ref{sec:vof}. The sign distance function is separately advected using a second order accurate Bell-Colella-Glaz \cite{bell1989second} scheme to obtain a provisional sign distance function $\phi^*$ as
\begin{equation}
    \phi^* = \phi^n + \Delta t \boldsymbol{\nabla} \cdot (\mathbf{u}^n \cdot \phi^n),
\end{equation}
where $\Delta t$ is the time step. The advection step is then followed by a PLIC reconstruction in each interfacial cell (where $0<C<1$) to obtain $\alpha$ and $\mathbf{m}$ from equation \eqref{eq:alpha} for a given value of the color function. Note that $\phi^{VOF} = \alpha \Delta$ ($\Delta$ being the grid size) is an approximation of the distance function calculated from VOF. We use $\phi^{VOF}$ to couple the level set and VOF methods as
\begin{equation}
    \phi^{**} = W \phi^{VOF} + (1 - W) \phi^*
    \label{eq:wtcoupling}
\end{equation}
where $\phi^{**}$ is yet another approximation of the sign distance function, and $W$ is a small {\color{blue} system-dependent} weight ($W = 0.1$, unless otherwise mentioned). Equation \eqref{eq:wtcoupling} corresponds to an exponential filtering of the level set distance function in time, i.e. an exponential relaxation of the level set distance function $\phi$ towards the VOF distance function $\phi^{VOF}$, on a timescale equal to $\Delta t/W$. {\color{blue}This exponential filtering ensures that in the stationary limit the distance function is equal to the exact (VOF) distance function in cells containing the interface, while minimizing the transient curvature noise induced by numerical errors linked to VOF advection (the weight parameter $W < 0.5$ ).}


The level set function calculated from equation \eqref{eq:wtcoupling} can be distorted and can also lose the signed distance property over a few time steps. Since this function will be used to calculate the surface tension stress tensor, we need a smooth function with the same position of
the zero level set. To obtain such a function, we employ the redistancing algorithm implemented by Limare et al. (see Appendix A in reference \cite{limare2023hybrid}) to obtain the sign distance function at the $(n+1)^{th}$ time step $\phi^{n+1}$. 
Once $\phi^{n+1}$ is calculated, we use it to compute the curvature as

\begin{equation}
    \kappa_{i,j} =\left(\boldsymbol{\nabla} \cdot \frac{\boldsymbol{\nabla} \phi}{|\boldsymbol{\nabla}\phi|}\right)_{i,j} = \left(\frac{\phi_x^2 \phi_{yy} - 2 \phi_x \phi_y \phi_{xy} + \phi_y^2 \phi_{xx}}{(\phi_x^2 + \phi_y^2)^{3/2}}\right)_{i,j},
\end{equation} 
where the derivatives are represented with subsripts ($\phi_x,\phi_{xx},\phi_{xy}$). These are calculated using a central difference scheme
\begin{align*}
    (\phi_x)_{i,j} &\approx \frac{\phi_{i+1,j} - \phi_{i-1,j}}{2 \Delta}, \\
    (\phi_{xx})_{i,j} &\approx \frac{\phi_{i+1,j} - 2 \phi_{i,j} + \phi_{i-1,j}}{\Delta^2}, \\
    (\phi_{xy})_{i,j} &\approx \frac{\phi_{i+1,j+1} - \phi_{i-1,j+1} - \phi_{i+1,j-1} + \phi_{i-1,j-1}}{4\Delta^2},
\end{align*}
tangent vectors 
\begin{equation}
    \mathbf{t} = \left[\phi_y, \phi_x \right] \approx \left[(\phi_{i,j+1/2} - \phi_{i,j-1/2})/\Delta , (\phi_{i+1,j} - \phi_{i-1,j})/\Delta \right]
\end{equation}
where $i,j$ represent the cell indices. Curvature $\kappa$ and tangent vector $\mathbf{t}$ are then used to calculate the surface tension stress tensor using algorithms \ref{alg:sigxx} and \ref{alg:sigxy} for $d = \phi^{n+1}$. Finally, the surface tension force is determined as the divergence of this stress tensor (equation \eqref{eq:ffromsig}) and then added to the momentum equation as a source term.

\subsubsection{Height function method (HF)}

\begin{figure}
    \centering
    \includegraphics{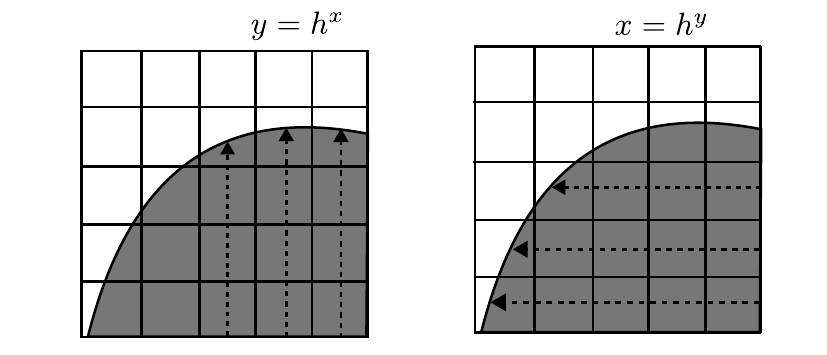}
    \caption{Representation of an interface with in a computational grid with height functions, shown with dotted arrows in $(a)$ Vertical direction $h^x$ and $(b)$ Horizontal direction $h^y$.}
    \label{fig:heights}
\end{figure}

The derivatives of interfacial shapes (normal, curvature, etc.) calculated directly from the color function $C$ can be inaccurate since the latter is not smooth. Therefore, smoothing operators are frequently applied prior to derivative calculations; however, even then, the resulting derivatives may still suffer from inaccuracies \cite{cummins2005estimating}. The height function method was proposed as an alternative to calculate a continuous representation of the interface from the color function and thus to calculate the derivatives of the interfacial shapes with greater accuracy \cite{helmsen1997non,sussman2003second}. It has been shown that machine accurate results for parasitic currents could be obtained upon combining the height function curvature computation with a well balanced method \cite{popinet2009accurate}. Since the HF method has several advantages, we propose another version of the integral formulation based on it, described in algorithm \ref{alg:hf} and discussed next. 
\begin{algorithm}
    \caption{Algorithm for HF method} \label{alg:hf}
    \begin{algorithmic}
    \Require $C^{n+1}$
    \State Calculate $h^x$ and $h^y$ from $C$ using the algorithm from reference \cite{popinet2009accurate}
    \For{Interfacial cell}
        \State $\mathbf{n} \gets \boldsymbol{\nabla} C/\vert \boldsymbol{\nabla} C\vert$
        \If{$n^x < 0.5$}
        \State $\kappa \gets \kappa(h^x), \mathbf{t} \gets \mathbf{t} (h^x)$
        \Else
        \State $\kappa \gets \kappa(h^y), \mathbf{t} \gets \mathbf{t} (h^y)$
        \EndIf
        \State Get weights $w^x,w^y$ from $n^x,n^y$ using equation \eqref{eq:weights}
        \State Obtain diagonal terms $\sigma^{xx}(d=h^x)$ and $\sigma^{xx}(d=h^y)$ from algorithm \ref{alg:sigxx}
        \State $\sigma^{xx} \gets w^x \sigma^{xx}(d=h^x) + w^y \sigma^{xx}(d=h^y)$            
    \EndFor
    \For{Interfacial Vertex}
        \State $\mathbf{n} \gets \boldsymbol{\nabla} C/\vert \boldsymbol{\nabla} C\vert$
        \If{$n^x < 0.5$}
        \State $\mathbf{t} \gets \mathbf{t} (h^x)$
        \Else
        \State $\mathbf{t} \gets \mathbf{t} (h^y)$
        \EndIf
        \State Get weights $w^x,w^y$ from $n^x,n^y$ using equation \eqref{eq:weights}
        \State Obtain non-diagonal terms $\sigma^{xy}(d=h^x)$ and $\sigma^{xy}(d=h^y)$ from algorithm \ref{alg:sigxy}
        \State $\sigma^{xy} \gets w^x \sigma^{xy}(d=h^x) + w^y \sigma^{xy}(d=h^y)$            
    \EndFor
    \State Calculate $F_\gamma$ using equation \eqref{eq:ffromsig} and add to Navier-Stokes momentum equation.
    \end{algorithmic}
\end{algorithm}

The height function field is a vector field $\mathbf{h} = \{h^x, h^y\}$ that is obtained by integrating the color function in respective directions on a stencil local to the interface
\begin{equation}
    h^x_{i,j} = \sum_{j-4}^{j} C_{i,j} \Delta,
\end{equation}
$h^y$ can be obtained similarly by rotation of indices $\{i,j\}$ (see figure \ref{fig:heights}). In Basilisk, height functions are evaluated on locally varying stencils following the algorithm detailed in reference \cite{popinet2009accurate}. After obtaining the discrete approximation of the interface height $y = h^x$ and/or $x = h^y$ similar to figure \ref{fig:heights}, we calculate the curvature as
\begin{equation}
    \kappa = \frac{h^{\prime \prime}}{(1 + h^{\prime2})^{3/2}},
\end{equation}
and the tangent vector as
\begin{equation}
    \mathbf{t} = \left[ \frac{h^\prime}{\sqrt{1 + h^{\prime 2}}}, -\frac{1}{\sqrt{1 + h^{\prime 2}}} \right]^{\rm{T}},
\end{equation}
where $h^\prime$ and $h^{\prime \prime}$ are the first and second order derivatives of $h$ that are calculated using a central difference scheme for example
\begin{equation}
    (h^x)^\prime \approx \frac{h^x_{i+1,j} - h^x_{i-1,j}}{2 \Delta}; (h^x)^{\prime \prime} \approx \frac{h^x_{i+1,j} - 2 h^x_{i,j} + h^x_{i,j+1}}{\Delta^2},
    \label{eq:calcprimes}
\end{equation}
where $i,j$ are cell indices. Depending on the orientation of the normal vector estimated from the color function $\mathbf{n} = \nabla C/|\nabla C|$, the interface is either represented as $y = h^x$ (when $|n^y| > |n^x|$), or $x = h^y$ (when $|n^x| > |n^y|$) to calculate the curvature and tangent vector. The height function in a cell also yields the distance from the cell center to the nearest interface in a particular direction, hence we can use it directly to obtain the diagonal terms $(\sigma^{xx} (d=h^x)$ and/or $\sigma^{xx} (d=h^y))$ and non-diagonal terms $(\sigma^{xy} (d=h^x)$ and/or $\sigma^{xy} (d=h^y))$ in the surface tension stress tensor using algorithms \ref{alg:sigxx} and \ref{alg:sigxy}, respectively. We use a weighted average of these terms to obtain a smooth estimate of each diagonal and non-diagonal term. The weights are calculated from the components of the normal $\left[n^x,n^y \right]^{\rm{T}}$ previously obtained from the color function 
\begin{equation}
    w^x = \begin{cases}
        0 ~~ \textrm{ if } (n^x)^2 < 0.15\\
        1 ~~ \textrm{ if } (n^x)^2 > 0.85 \\
        (n^{x})^{2} ~~\textrm{Otherwise}
    \end{cases}, ~~ w^y = \begin{cases}
        0 ~~ \textrm{ if } (n^y)^2 < 0.15\\
        1 ~~ \textrm{ if } (n^y)^2 > 0.85 \\
        (n^{y})^{2} ~~\textrm{Otherwise}      
    \end{cases},
    \label{eq:weights}
\end{equation}
then, each of the diagonal and non-diagonal terms are obtained by weighting terms calculated from $h^x$ and $h^y$ with $w^x$ and $w^y$ respectively as
\begin{eqnarray}
    \sigma^{xx} = w^x \sigma^{xx} (d=h^x) + w^y \sigma^{xx}(d=h^y),\\
    \sigma^{xy} = w^x \sigma^{xy} (d=h^x) + w^y \sigma^{xy}(d=h^y).
    \label{wtsum}
\end{eqnarray}
\noindent Finally, the surface tension force is estimated from the divergence of this calculated stress tensor and is added to the momentum equation.




\subsubsection{Height function to distance method (HF2D)}
The HF2D method is described in algorithm \ref{alg:hf2d}. In this method, we obtain the distance function from the height functions. We construct the height functions field similarly to the HF method. After constructing the height functions, we fit a parabolic curve to represent the interface
\begin{equation}
    y(x) = h^{\prime \prime} x^2 + h^\prime x + h,
    \label{eq:parabola}
\end{equation}
using the horizontal height functions $h = h^x$, for example the solid curve in figure \ref{fig:heights}$a$. The fitted curve $x(y)$ similar to the example in figure \ref{fig:heights}$b$ is also fitted from the vertical height functions $h = h^y$. An accurate prediction of the sign distance function $d$ is then obtained by calculating the minimum distance from the center of the cell to the parabolic curve (equation \eqref{eq:parabola}). In the cases where we could obtain two estimates of the sign distance function $d(x)$ and $d(y)$ from both $h^x$ and $h^y$, respectively, we take an average of both; otherwise, we use only $d(x)$ or $d(y)$ depending on availability. This sign distance function is then used to compute the terms of the surface tension stress tensor and, therefore, the surface tension force using the algorithm \ref{alg:sigxx} and \ref{alg:sigxy}, which is subsequently added to the momentum equation.

\begin{algorithm} [htb!]
\caption{Algorithm for HF2D method}\label{alg:hf2d}
\begin{algorithmic}
\Require $C^{n+1}$

\State Calculate $h^x$ and $h^y$ from $C$ using the algorithm from reference \cite{popinet2009accurate}
\For{$x_i \in \{x,y\}$}
\State $a_{x_i} \gets  (h^{x_i})^{\prime \prime}$, $b_{x_i} \gets  (h^{x_i})^\prime$, $c_{x_i} \gets h^{x_i},$  calculated using equation \eqref{eq:calcprimes}
\State $y(x_i) = a_{x_i} x_i^2 + b_{x_i} x_i + c_{x_i}$
\State Calculate the minimum distance $d(x_i)$ from the parabola $y(x_i)$ to the cell center 
\EndFor
\If{ Both $d(x)$ and $d(y)$ are available}
\State $d \gets \frac{d(x) + d(y)}{2}$ 
\Else \State $d \gets d(x)$ or $d \gets d(y)$, whichever is available
\EndIf

\State $\kappa \gets \kappa(d), t^x \gets t^x (d)$
\For{All cells}
\State Calculate $\sigma^{xx} (d)$ and $\sigma^{yy} (d)$ using Algorithm \ref{alg:sigxx}
\EndFor
\For{All vertices}
\State Calculate $\sigma^{xy} (d)$ and $\sigma^{yx} (d)$ using Algorithm \ref{alg:sigxy}
\EndFor
\State Calculate $F_\gamma$ using equation \eqref{eq:ffromsig} and add to Navier-Stokes momentum equation.
\end{algorithmic}
\end{algorithm}

\subsection{Navier-Stokes solver}
The Navier-Stokes solver which is already implemented in Basilisk is used for the present study. The source code is available at \url{http://basilisk.fr/src/navier-stokes/centered.h} and the method is described in detail in references \cite{popinet2003gerris,popinet2009accurate,lagree2011granular}. Here, we only briefly discuss the equations and the solution procedure. We use a time split projection method \cite{chorin1969convergence} which is second order accurate and staggered in time giving the semi-discrete equations
\begin{equation}
    \centering
    \frac{C^{n + \frac{1}{2}} - C^{n-\frac{1}{2}}}{\Delta t} + \boldsymbol{\nabla} \cdot (\mathbf{u}^n C^n) = 0,
    \label{eq:colorfndis}
\end{equation}
\begin{equation}
\centering
    \rho^{n + \frac{1}{2}} \left(\frac{\mathbf{u}^* - \mathbf{u}^n}{\Delta t} + \mathbf{u}^{n+\frac{1}{2}} \cdot \boldsymbol{\nabla} \mathbf{u}^{n+\frac{1}{2}}\right) = \boldsymbol{\nabla} \cdot \left[\mu^{n+\frac{1}{2}} (\mathbf{D}^{n} + \mathbf{D}^{*})\right] + F_\gamma^{n+\frac{1}{2}},
    \label{eq:discretemom}
\end{equation}
\begin{equation}
    \mathbf{u}^{n+1} = \mathbf{u}^* - \frac{\Delta t}{\rho^{n+\frac{1}{2}}} \boldsymbol{\nabla} p^{n+\frac{1}{2}},
    \label{eq:un+1}
\end{equation}
\begin{equation}
    \boldsymbol{\nabla} \cdot \mathbf{u}^{n+1} = 0,
    \label{eq:divv}
\end{equation}
where the superscripts $n,n+\frac{1}{2}$ and $n+1$ refer to the time coordinates $t = t^n$, $t = t^{n+\frac{1}{2}}$ (provisional), and $t^{n+1} = t^n + \Delta t$. 

Now, we briefly describe the algorithm to solve equations \eqref{eq:colorfndis}-\eqref{eq:divv}. The first step is the advection of the color function (equation \eqref{eq:colorfndis}) using the direction split algorithm of Weymouth and Yue \cite{weymouth2010conservative}, also described in Section \ref{sec:vof}. Then, the flux for the advection term $\mathbf{u}^{n+\frac{1}{2}} \cdot \boldsymbol{\nabla} \mathbf{u}^{n+\frac{1}{2}}$ in the momentum equation is calculated using the second-order Bell-Colella-Glaz scheme \cite{bell1989second}. After the advection step, we calculate $\mathbf{F}_\gamma^{n+\frac{1}{2}}$ following the discussions from section \ref{sec:surfacetension}. Next, we obtain a temporary velocity field $\mathbf{u}^*$ by solving the rearranged form of equation \eqref{eq:discretemom} written as
\begin{equation}
    \frac{\rho^{n+\frac{1}{2}}}{\Delta t} \mathbf{u}^* - \boldsymbol{\nabla} \cdot \left(\mu^{n+\frac{1}{2}} \mathbf{D}^*\right) = \boldsymbol{\nabla} \cdot \left(\mu^{n+\frac{1}{2}} \mathbf{D}^{n}\right) + \mathbf{F}_\gamma^{n+\frac{1}{2}} + \rho^{n+\frac{1}{2}} \left(\frac{\mathbf{u}^n}{\Delta t} - \mathbf{u}^{n+\frac{1}{2}} \cdot \boldsymbol{\nabla} \mathbf{u}^{n+\frac{1}{2}}\right),
    \label{eq:momvisc}
\end{equation}
using a version of multigrid solver implemented in Basilisk described in detail in reference \cite{popinet2003gerris,popinet2009accurate,saade2023multigrid}. All the terms on the right hand side of equation \eqref{eq:momvisc} have already been calculated previously, therefore, these terms are simply added as a source term to the multigrid solver. We use a second-order accurate central difference scheme for calculating the velocity derivatives required for $\mathbf{D^*}$ and a second order accurate Crank-Nicholson discretization of the viscous terms, thus ensuring a second order accurate method in both space and time. Next, we calculate the pressure field $p^{n+\frac{1}{2}}$ by solving a Poisson equation resulting from the divergence free condition for the velocity field $(\nabla \cdot \mathbf{u}^{n+1} = 0)$. The Poisson equation is obtained by taking the divergence of equation \eqref{eq:un+1} and equating it to zero that results in the following equation
\begin{equation}
    \boldsymbol{\nabla} \cdot \left[\frac{\Delta t}{\rho^{n+\frac{1}{2}}} \boldsymbol{\nabla} p^{n+\frac{1}{2}}\right] = \boldsymbol{\nabla} \cdot \mathbf{u}^*.
\end{equation}
This equation is solved by an iterative solver that runs until the relative error in each cell falls below $10^{-6}$. We use a collocated arrangement of variables such that both pressure and velocity are defined at the cell centers. To avoid the famous checkerboard problem of collocated variables, we use an approximate projection method for the spatial discretization of the Poisson equation for pressure \cite{popinet2003gerris,popinet2009accurate}. Finally, using the calculated pressure field $p^{n+\frac{1}{2}}$ and the temporary velocity field $ \mathbf{u}^*$, we compute $\mathbf{u}^{n+1}$ from equation \eqref{eq:un+1}.


\section{Test cases \label{sec:Validation}}
\subsection{Damping of a capillary wave\label{sec:damping}}
\begin{figure}[hb!]
    \centering
    \includegraphics[scale = 0.8]{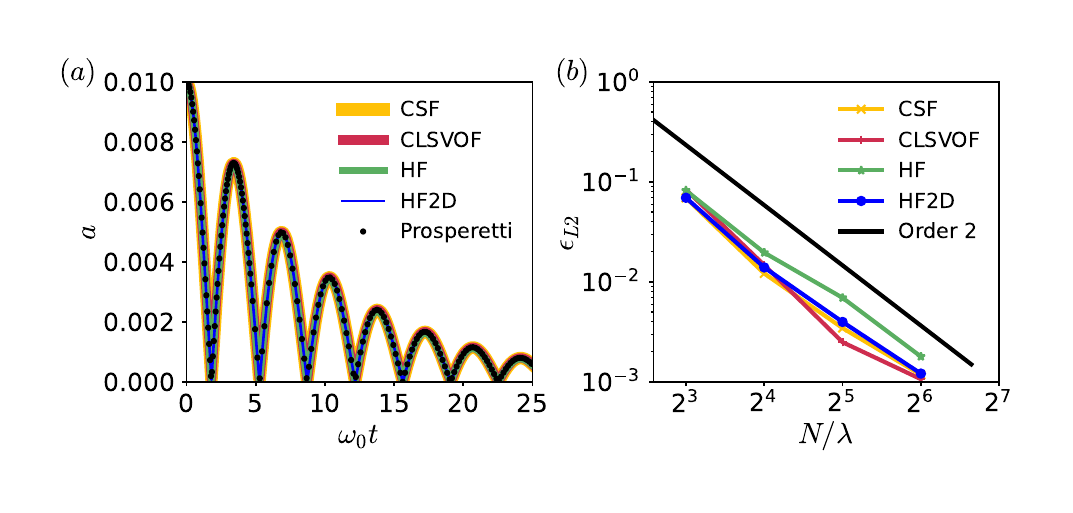}
    \caption{Capillary wave damping test case for all three methods CLSVOF, HF, HF2D {\color{red} for a fixed Laplace number $\rm{La} = 6000$} $(a)$ The evolution of the maximum perturbation amplitude. $(b)$ Convergence of the $L2$ norm of the numerical errors with the number of grid points per wavelength of the capillary wave $N/\lambda$. The code to reproduce these results is available at \url{http://basilisk.fr/sandbox/msaini/Marangoni/capwave.c}.}
    \label{fig:capwave}
\end{figure}
Capillary waves are often generated in flows driven by surface tension and are damped through viscous dissipation. For sufficiently small amplitudes, this damping can be predicted by solving an initial value problem, as shown by Prosperetti \cite{prosperetti1981motion}. 
A standard test case based on this solution was proposed for the validation of the surface tension formulation by Popinet and Zaleski \cite{popinet1999front}. We consider a flat 2D interface between two identical fluids at rest. This interface, located at the center of the numerical domain, is perturbed with a cosine function of wavelength $\lambda = 2 \textrm{ units}$ and amplitude $a = 0.01 \lambda$. The dynamics is controlled by dimensionless Laplace number given as $$ \textrm{La} = \frac{\gamma \rho \lambda}{\mu^2} = 6000.$$ The numerical domain is a square of size $\lambda$, with periodic left and right boundaries, and free-slip top and bottom boundaries. 
The evolution of the maximum amplitude of the perturbation $a_{max}$ is shown in figure \ref{fig:capwave}$a$ for all three methods described previously. Good agreement is achieved with the theoretical predictions of Prosperetti \cite{prosperetti1981motion}. The error is evaluated with the L2 norm 
\begin{equation}
    \epsilon_{\textrm{L2}} = \frac{1}{\lambda} \sqrt{ \frac{\omega_0}{25} \sum_{t=0}^{25/\omega_0} (a_{max} - a_{max}^{exact})^2}
\end{equation}
where $a_{max}^{exact}$ is the theoretical estimation of the maximum amplitude, and $\omega_0$ is the frequency of the normal mode given as 
\begin{equation}
\omega_0^2 = \frac{\gamma}{2 \rho} \left(\frac{2 \pi}{\lambda}\right)^3,
\end{equation}
the error is also shown in figure \ref{fig:capwave}$b$, where we show that all methods have a second order convergence with the refinement of the grid size for this test case.

\subsection{Spurious currents around a translating interface}
This test case was proposed by Popinet \cite{popinet2009accurate} and was discussed in detail by Abadie \textit{et al.} \cite{abadie2015combined}. In this test case, a circular interface is placed in a uniform flow with horizontal velocity $U$. The periodic boundary condition is prescribed on both the left and right boundaries, and the symmetry boundary condition is imposed on both the top and bottom boundaries. Both fluids, i.e. inside and outside the interface, are assumed to be identical, so that there is no contrast in their viscosity and density. In such a case, the interface also moves with a velocity $U$ and in the reference frame of the moving interface, Laplace's law predicts the jump in pressure across the interface. The pressure $p_i$ inside a circular interface of radius $R$ is higher than that of the outside pressure $p_o$ due to surface tension $\gamma$
\begin{equation}
    p_i = p_o + \frac{2\gamma}{R}.
\end{equation}
 Obeying Laplace's law down to machine accuracy in each grid cell is a challenge, and numerical errors can result in the emergence of spurious velocity currents near the interface. Numerical errors arise during the advection of the interface given by equation (\eqref{eq:colorfn}). These errors lead to small distortions in the color function field, which can subsequently cause numerical errors in the curvature calculation. As a result, fluctuations appear in the pressure field near the interface, leading to spurious oscillations in the velocity field. 
Here, we consider a circular interface of diameter $D = 0.4$ dimensionless units in a domain that is a square box of length $L = 1$ units. 
Three time scales appear in this problem, i.e. the viscous time scale $$t_{\mu} = \frac{\rho D^2}{\mu},$$ indicating the time it takes for momentum to diffuse across the diameter, the capillary time scale $t_\gamma$ associated with the propagation speed of capillary waves $$t_\gamma = \sqrt{\frac{\rho D^3}{\gamma}},$$ and another time scale $t_U$ associated with the background velocity field, $$t_U = \frac{D}{U}.$$ The problem can be characterized by two dimensionless numbers, namely the Laplace number $$\textrm{La} = \left(\frac{t_\mu}{t_\gamma}\right)^2 = \frac{\rho \gamma D}{\mu^2},$$ and the Weber number $$\textrm{We} = \left(\frac{t_\gamma}{t_U}\right)^2 = \frac{\rho U^2 D}{\gamma}.$$

\begin{figure}[hbt!]
    \centering
    \includegraphics[scale = 0.6]{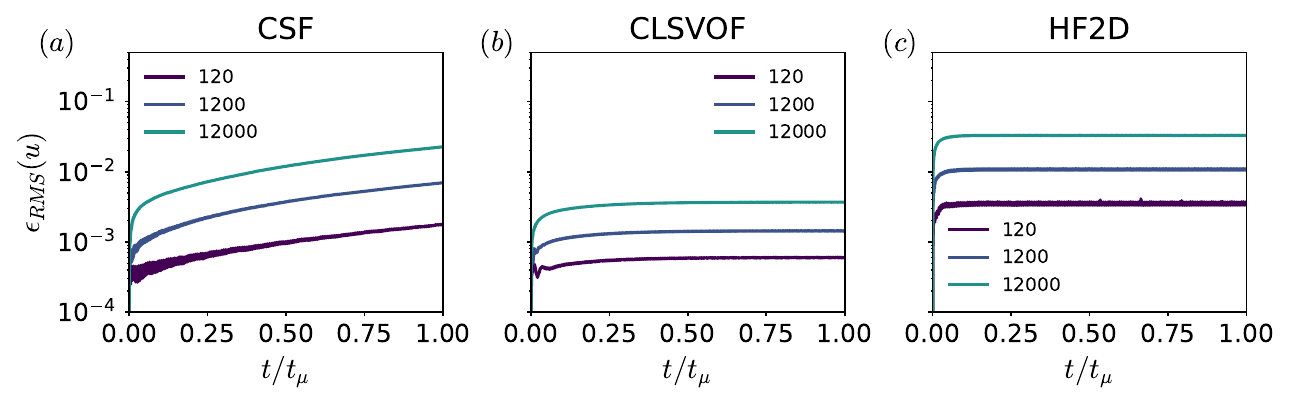}
    \caption{
    Evolution of the RMS value of spurious velocity currents for a translating circular interface {\color{red} for a fixed Weber number $\rm{We} = 0.4$ and a varying Laplace number $\rm{La} = 120, 1200, 12000$} with $(a)$ the standard CSF method of Basilisk $(b)$ the CLSVOF method $(c)$ the HF2D method. The code to reproduce these results is available at \url{http://basilisk.fr/sandbox/msaini/Marangoni/spuriousMov.c}.}
    \label{fig:spurious}
\end{figure}

\begin{figure}[hbt!]
    \centering
    \includegraphics[scale = 0.9]{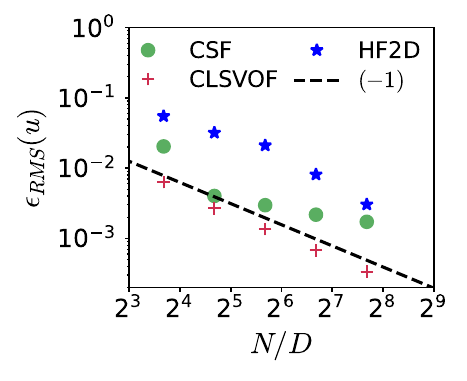}
    \caption{Convergence of RMS value of the spurious velocity oscillations for the standard CSF method of Basilisk, the CLSVOF method, and the HF2D method for a fixed Weber number $\textrm{We} = 0.4$ and a fixed Laplace number $\textrm{La} = 12000$. A dashed line with a slope of $-1$ is also shown.}
    \label{fig:spuriousConv}
\end{figure}

We have analyzed the time evolution of spurious velocity oscillations in the translating frame of reference for three Laplace numbers $\textrm{La} = 120,1200,12000$ at a fixed Weber number $\textrm{We} = 0.4$. In figure \ref{fig:spurious}, we show the time evolution of the root mean square (RMS) value of the spurious velocity oscillations 
\begin{equation}
    \epsilon_{RMS}(u) =\frac{1}{U} \sqrt{\frac{\sum_{j} \left(u - U\right)^2 \Delta_j^2}{\sum_j \Delta_j^2}},
\end{equation}
where $j$ is the index of the computational cell and $\Delta_j$ is its size. For this case, we do not discuss the results from the HF method because the spurious currents in the HF method method grow quickly, causing the simulations to blow up. The results of the following three methods are reported: a) the standard CSF method with VOF \cite{popinet2009accurate,popinet2003gerris}, b) the newly developed integral formulation with CLSVOF and c) HF2D.  In all the cases, the spurious velocity currents do not decay to machine accuracy within the viscous time scale because the errors from the advection of the interface continuously introduce perturbations to the interface's shape and thus to the curvature which then cause the spurious velocity oscillations. Remarkably, the smallest oscillations among the three methods are seen in the CLSVOF method. This could be attributed to the fact that in CLSVOF the errors are smoothed over time as the sign distance function (used for the curvature calculation) relaxes to the VOF distance function slowly over 10 time steps as a result of the weight $W = 0.1$ in equation \eqref{eq:wtcoupling}. It should also be noted that unlike CLSVOF and HF2D, errors in the CSF method continuously increase until $t = t_\mu$.

\begin{figure}[h!]
    \centering
    \includegraphics[scale = 0.8]{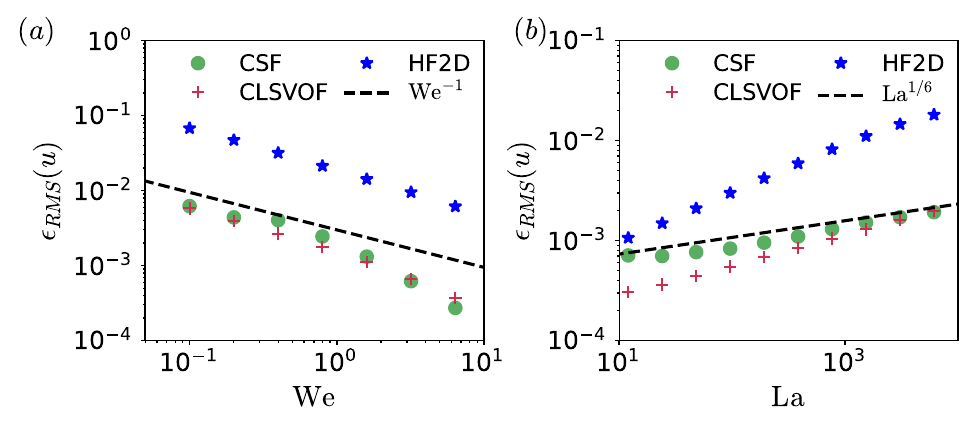}
    \caption{Scaling of spurious oscillations for the standard CSF method of Basilisk, the CLSVOF method, and the HF2D method $(a)$ with the Weber numbers at a fixed Laplace number of $\textrm{La} = 12000$. A dashed line with a slope of $-1$ is also shown. $(b)$ with the Laplace number at a fixed Weber number $\textrm{We} = 0.4.$ A dashed line with a slope of $1/6$ is also shown, consistent with reference \cite{popinet2009accurate}.}
    \label{fig:spuriousscaling}
\end{figure}

Figure \ref{fig:spuriousConv} shows the convergence of the velocity errors for the three methods for the Laplace number $\textrm{La} = 12000$, Weber number $\textrm{We} = 0.4$ at time $t = 2.5 t_U$. Both CLSVOF and HF2D show close to first-order convergence, however, the errors in the CSF method show significantly slower convergence. The scaling of the error with the Weber and Laplace numbers at time $t = 2.5 t_U$ is shown in figure \ref{fig:spuriousscaling}. As the Weber number increases at a fixed Laplace number $\textrm{La} = 12000$, the surface tension force decreases, leading to a decrease in the velocity errors. For small Weber numbers, the errors scale as $\textrm{We}^{-1}$ for all three methods. For both CSF and CLSVOF, the error decays faster than $\textrm{We}^{-1}$ for Weber numbers larger than 1. Alternatively, when the Laplace number is varied at fixed Weber number $\textrm{We} = 0.4$, the velocity errors behave differently for all three methods. The spurious oscillations decrease upon decreasing the Laplace number because of increased viscous dissipation. For the CSF method, a scaling of $\textrm{La}^{1/6}$ is observed which is consistent with the results of \cite{popinet2009accurate}. However, errors scale with an exponent greater than $1/6$ for both the CLSVOF and HF2D methods.

\subsection{Rising bubble}
\begin{figure}[h!]
    \centering
    \includegraphics[scale = 0.85]{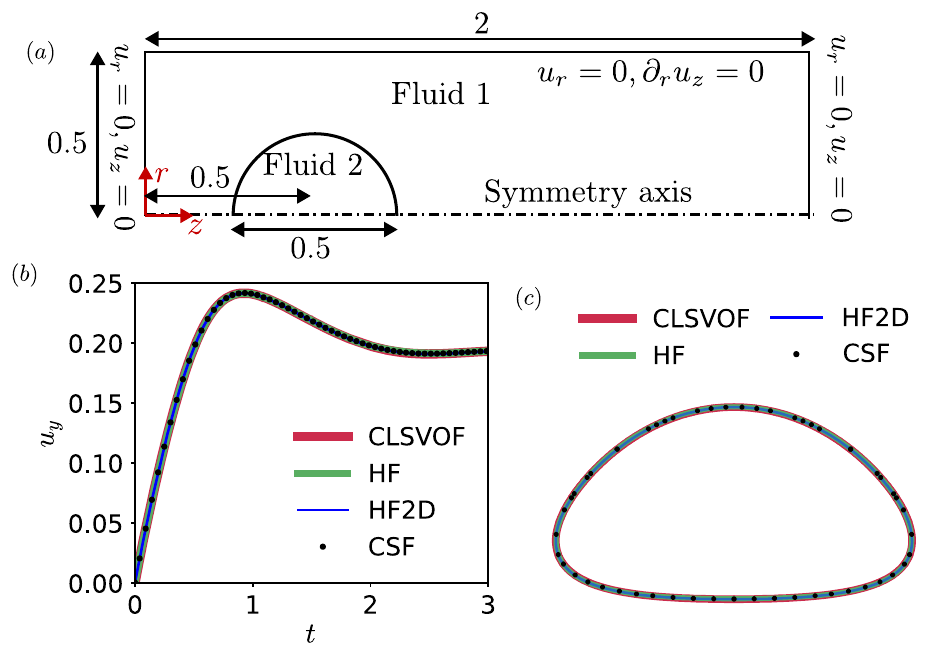}
    \caption{$(a)$ Numerical setup for the rising bubble test case as proposed by Hysing et al. \cite{hysing2009quantitative} {\color{red} for Galilei number $\rm{Ga} = 2.25$, and E\"{o}tv\"{o}os number $\rm{Eo} = 124$}. $(b)$ Rise velocity of the bubble obtained from the four methods. $(c)$ The shape of the bubbles at 3 time units for the four methods. The code to reproduce the results in this figure is available at \url{http://basilisk.fr/sandbox/msaini/Marangoni/rising.c}.}
    \label{fig:risingbub}
\end{figure}

A bubble rising in a quiescent liquid has a very rich phenomenology due to the interplay between surface tension, viscosity, and gravity \cite{tripathi2015dynamics}. A quantitative benchmark based on the problem of a rising bubble was proposed by Hysing \textit{et al.} \cite{hysing2009quantitative} for which the numerical setup is shown in figure \ref{fig:risingbub}$a$. We assume a bubble of radius 0.25 dimensionless units initially kept in a tube of radius 0.5 units and length 2 units, and at a distance of 0.5 units from the left boundary as shown in figure \ref{fig:risingbub}$a$. The boundary condition for the bottom boundary is symmetry, the top boundary obeys a homogeneous Dirichlet condition for the radial velocity and a homogeneous Neumann condition for the axial velocity; for both the left and the right boundaries, a homogeneous Dirichlet condition is applied for the radial as well as the axial velocity. Both the density and viscosity ratios of the bubble relative to its surrounding fluid are 1:10. 
The problem can be defined with the Galilei number $$\textrm{Ga} = \frac{(\rho_1 - \rho_2)^2 g D^3}{\mu_1^2} = 2.25,$$ and the E$\ddot{\text{o}}$tv$\ddot{\text{o}}$s number $$ \textrm{Eo} = \frac{(\rho_1 - \rho_2) D^2 g}{\gamma} = 124,$$ where $\rho_1$ and $\rho_2$ are the densities of the outer fluid and of the bubble, respectively, $D$ is the diameter of the bubble, $\mu_1$ is the viscosity of the outer fluid, $g$ is the acceleration due to gravity, and $\gamma$ is the surface tension. The rise velocity as a function of time, and the shape of the bubbles at 3 time units are shown in figures \ref{fig:risingbub}$b$ and \ref{fig:risingbub}$c$, respectively. For both the bubble shape and the rise velocity, the numerical predictions of all methods (CLSVOF, HF, and HF2D) match well with the expected result obtained while using the standard CSF method of Basilisk.

\subsection{Marangoni induced translation of a drop \label{sec:ygb}}
We now consider flows that are governed by the variation of surface tension along the interface. We consider a classical problem of translating drop due to a surface tension gradient. Young \textit{ et al.} \cite{young1959motion} studied a similar problem in detail and provided an analytical solution for the limiting case of a creeping flow and a large thermal conductivity of the fluids. We assume that the surface tension $\gamma$ varies linearly with the local average temperature $T$ as
\begin{equation}
\gamma = \gamma_0 + \gamma_T (T - T_0),    
\end{equation}
where $\gamma_0$ is the surface tension at the reference temperature $T_0$ and $\gamma_T$ is a coefficient for linear fit. The temperature $T$ obeys an advection-diffusion equation 
\begin{equation}
    \partial_t T + \mathbf{u} \cdot \nabla T = \nabla \cdot \left(\alpha \nabla T\right),
    \label{eq:advdiffT}
\end{equation} 
with the local fluid velocity $\mathbf{u}$, and the average thermal diffusivity $\alpha = C \alpha_1 + (1 - C) \alpha_2$ where $\alpha_1$ and $\alpha_2$ are the thermal diffusivities of the bulk and drop respectively. The characteristic length scale for the problem is the radius of the drop $R$, the density scale is the density of the bulk fluid $\rho_1$, the velocity scale is $ U = R \gamma_T \nabla T/\mu_1 $ where $\mu_1$ is the viscosity of the bulk fluid, and the temperature scale is $R \nabla T$. This leads to three independent dimensionless quantities, the Marangoni number $$\textrm{Ma} = UR/\alpha_1,$$ the Reynolds number $$\textrm{Re} = \rho_1 UR/\mu_1,$$ and the Capillary number $$\textrm{Ca}=\mu_1 U/\gamma_0.$$ The analytical solution for the translation velocity obtained by Young \textit{et al.} \cite{young1959motion} is
\begin{equation}
    u_{YGB} = - \frac{2}{(2 + 3 \mu_{2}/\mu_{1}) (2 + \alpha_{2}/\alpha_{1})} \frac{\gamma_T R \nabla T}{\mu_{1}},
    \label{eq:ygb}
\end{equation}
valid in the limit of $\textrm{Ma} \ll 1$, $\textrm{Ca} \ll 1$ and $\textrm{Re} \ll 1$. To compare with Young's analytical solution, we solve a simpler problem described by Al Saud \textit{et al.} \cite{abu2018conservative}, for which $\textrm{Re} = 0.066$ and $\textrm{Ca} = 0.6$. For the first test, the problem is further simplified by assuming $\mu_1 = \mu_2 = 1$ and $\rho_1 = \rho_2$ = 1, while the effect of viscosity and diffusivity contrast across the interface is analyzed in subsequent tests. 
For the first case, it is also assumed that the temperature varies linearly in the direction $z$, which is a solution of equation \eqref{eq:advdiffT} in the limit of $\alpha \gg UR$ such that the Marangoni number is effectively zero. For subsequent cases where we vary the ratio of thermal diffusivities $\alpha_2/\alpha_1$, we solve equation \eqref{eq:advdiffT} with $\textrm{Ma} = 10^{-6}$.

\begin{figure}[h!]
    \centering
    \includegraphics[scale=0.85]{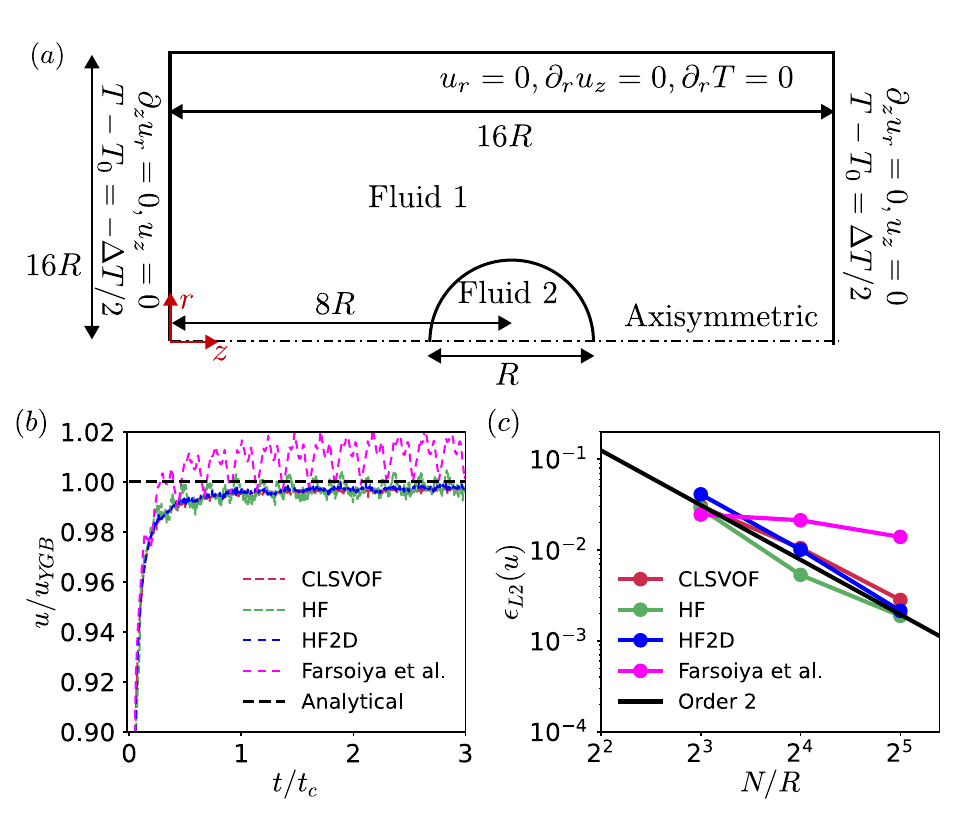}
    \caption{Marangoni induced translation of a drop {\color{red} for a linear surface tension gradient with Reynolds number $\rm{Re} = 0.066$ and Capillary number $\rm{Ca} = 0.6$.} $(a)$ Numerical setup for the case of a droplet translating due to a Marangoni flow $(b)$ Evolution of the drop velocity as compared to the prediction of Young \textit{et al.} \cite{young1959motion} i.e. equation \eqref{eq:ygb}. $(b)$ Comparison of the drop velocity to equation \eqref{eq:ygb} for the CLSVOF, the HF and the HF2D methods. The code to reproduce these results is available at \url{http://basilisk.fr/sandbox/msaini/Marangoni/marangoni.c}.}
    \label{fig:ygb}
\end{figure}

\begin{figure}[ht!]
    \centering
    \includegraphics{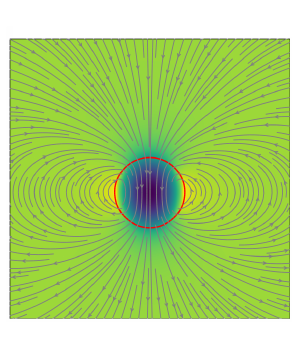}
    \caption{The velocity fields near the drop for Young's case obtained from the CLSVOF method at  $t = 3 U/R$ {\color{red} for a linear surface tension gradient with Reynolds number $\rm{Re} = 0.066$ and Capillary number $\rm{Ca} = 0.6$.}. The velocity field from the HF and the HF2D methods are analogous to that of the CLSVOF method.}
    \label{fig:vfields}
\end{figure}

The domain is 16 times the radius of the drop to minimize confinement effects and the numerical setup is shown in figure \ref{fig:ygb}$a$. Figure \ref{fig:ygb}$b$ shows the evolution of the average translation velocity of the drop $u = \sum_i u_z (1 - C_i) r_i \Delta_i^2/\sum_i(1 - C_i) r_i \Delta_i^2$, where $u_z$ is the local axial velocity in the cell with index $i$ and size $\Delta_i$, and $\sum_i(1 - C_i) r_i \Delta_i^2$ is the volume of fluid $2$ in that cell. For this case, we also show a comparison with the results obtained using the CSF method of Farsoiya et al. \cite{farsoiya2024coupled} with their code shared freely at the Basilisk website \url{http://basilisk.fr/sandbox/farsoiya/marangoni_surfactant/marangoni.h}. The translation velocity for all the methods agrees well with the theoretical predictions of equation \eqref{eq:ygb}. However, observable numerical oscillations appear in the velocity of the drop, particularly in the case of the HF method and the CSF method of Farsoiya et al.. These two methods mainly rely on the height functions, with oscillations potentially emerging from cells lacking access to vertical or horizontal height functions. 
It should be noted that for all three methods CLSVOF, HF and HF2D methods, the velocity predictions are much more accurate ($\sim 99.99\%$) compared to other continuum surface force implementations of the variable surface tension term ($\sim 90 - 96 \%$) \cite{ma2011direct,samareh2014thermocapillary,balcazar2016level,stricker2017numerical,seric2018direct,tripathi2015dynamics}.  The $L2$ norm of error 
\begin{equation}
    \epsilon_{L2}(u) = \frac{1}{u_{YGB}}\sqrt{ \frac{ \sum_{2.5 t_c}^{3 t_c} (u(t) - u_{YGB})^2 (\Delta t)^2}{\sum_{2.5 t_c}^{3 t_c} (\Delta t)^2}},
\end{equation}
where $u(t)$ is the average drop velocity at time $t$ and $\Delta t$ is the time step. For all three currently developed methods, we obtain a second-order convergence of the error with respect to the analytical predictions for this problem, as shown in figure \ref{fig:ygb}$c$. {\color{red} We believe that it is the first demonstration of second-order convergence for the important case of translation of a droplet due to Marangoni stresses.} the CSF method of Farsoiya et al. \cite{farsoiya2024coupled}, the solution appears to be converging to a slightly wrong value, that is $1.01 u_{YGB}$. In Figure \ref{fig:vfields}, we show the velocity field near the interface in the case of the CLSVOF method. The velocity fields obtained from all the methods are essentially the same, with only minor differences appearing because of small oscillations that are hardly noticeable.

To quantify the relative computational speed of each method, we also calculate the physical elapsed time during the simulation per total number of grid points and per total number of iterations during a simulation (see Table \ref{tab:time}). As expected, the elapsed time for the CLSVOF method is more than that consumed by the HF method, since more numerical operations are required to calculate the surface tension forces (for example, redistancing) in the former as compared to the latter. Additionally, the computational cost of the HF2D method is intermediate and the elapsed time lies between those of the CLSVOF and the HF methods. The CSF method of Farsoiya \textit{et al.} is less computationally expensive compared to all other methods. {\color{red} However, the total simulation time is larger for the CSF method due to the velocity fluctuations.}  A detailed discussion on the computational time elapsed in different parts of the solvers is shown in the appendix A.

\begin{table}[h!]
\centering
\begin{tabular}{|l|l|l|l|l|}
\hline
                 & CLSVOF & HF   & HF2D & Farsoiya et al. \cite{farsoiya2024coupled} \\ \hline
$\frac{\textrm{Time }(\mu s)}{\textrm{Points } \cdot \textrm{Steps}}$ & 3.98   & 3.39 & 3.47 & 2.92 \\ \hline
$\textrm{Time (s)}$ & 18.28   & 16.31  & 17 & 25 \\ \hline
\end{tabular}
\caption{Total time elapsed per grid point per iteration {\color{red} and total simulation time} during the simulations of the Young's case for the CLSVOF, the HF, and the HF2D methods for simulations ran on 1 core of AMD Ryzen 9 5900HX with 3.3 GHz max frequency and 16 Giga Bytes of random access memory.
}
\label{tab:time}
\end{table}

\begin{figure}
    \centering
    \includegraphics[scale=0.9]{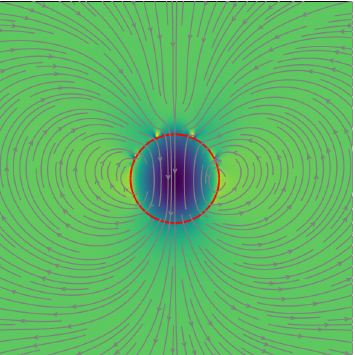}
    \caption{The velocity field around the drop driven by Marangoni flow as given by the HF method for $\mu_2/\mu_1 = 10$, {\color{red} Marangoni number $\rm{Ma} = 10^{-6}$, Reynolds number $\rm{Re} = 0.066$ and Capillary number $\rm{Ca} = 0.6$.} and at $t = 0.08 U/R$.}
    \label{fig:vfieldmu}
\end{figure}

\noindent \emph{Effect of viscosity and thermal diffusivity ratios} \\
Next, we consider the effect of varying both the viscosity ratio and the thermal diffusivity ratios between the drop and the bulk. As predicted by equation \eqref{eq:ygb}, the translation velocity decreases significantly when either the viscosity or thermal diffusivity ratio increases. The decay in velocity is predicted well with the CLSVOF and HF2D methods, while for the HF method there is a substantial deviation from the theoretical predictions at large viscosity and thermal diffusivity ratios due to numerical oscillations. It is important to observe that when the ratio of viscosity or thermal diffusivity is large, the drop's migration velocity decreases and may become comparable to the numerical oscillations encountered in the HF method. The careful diagnosis of the flow field at early times for such cases (for example, figure \ref{fig:vfieldmu}) signals that these oscillations originate from cells where either the vertical or horizontal height function is unavailable.

\begin{figure}[h!]
    \centering
    \includegraphics[scale=0.75]{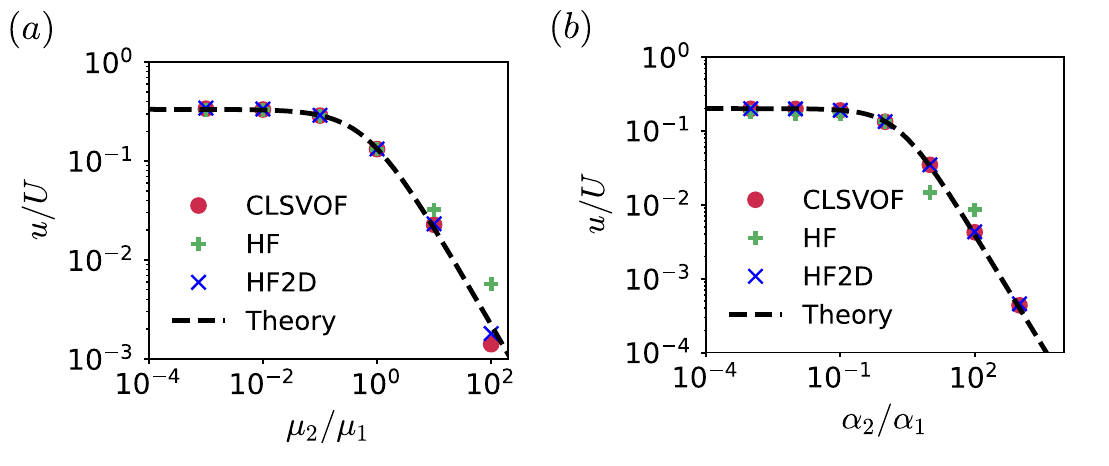}
    \caption{Effect of the $(a)$ Viscosity ratio $(b)$ Thermal diffusivity ratio on the drop migration velocity and comparison with the predictions of Young \textit{et al.} \cite{young1959motion}. The simulation parameters are {\color{red} Marangoni number $\rm{Ma} = 10^{-6}$, Reynolds number $\rm{Re} = 0.066$ and Capillary number $\rm{Ca} = 0.6$.}}
    \label{fig:ygbmu}
\end{figure}

\subsection{Interactions of a drop translating due to Marangoni flow with a plane wall}
\begin{figure}[h!]
    \centering
    \includegraphics[scale = 0.85]{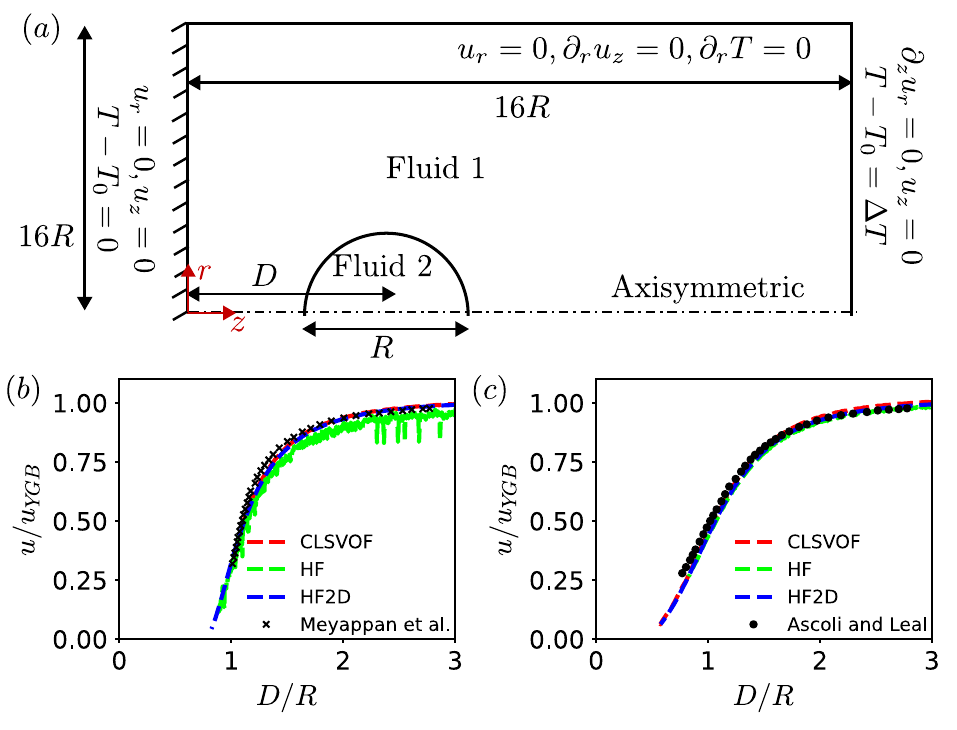}
    \caption{Interactions of drop translating due to Marangoni flow with a planar wall for {\color{red} Marangoni number $\rm{Ma} = 10^{-6}$, Reynolds number $\rm{Re} = 0.066$ and a varying Capillary number $\rm{Ca} = 0.15,4.55$}. $(a)$ Numerical setup 
    q$(b)$ The variation of the drop velocity with the distance from the wall for a Capillary number $\textrm{Ca} = 0.15$, for the CLSVOF, the HF and the HF2D methods, and a comparison with Meyappan \textit{et al.}'s results \cite{meyyappan1981thermocapillary}.  $(c)$  The variation of the drop velocity with the distance from the wall for a Capillary number $\textrm{Ca} = 4.55$, for the CLSVOF, the HF and the HF2D methods, and a comparison with Ascolli and Leal's results \cite{ascoli1990thermocapillary}. The code to reproduce the results in this figure is available at \url{http://basilisk.fr/sandbox/msaini/Marangoni/marangoniwall.c}.}
    \label{fig:deformable}
\end{figure}

A drop migrating due to the Marangoni flow normal to a plane wall slows down as it approaches this wall. The decrease in drop velocity was calculated by Meyyappan \textit{ et al.}. \cite{meyyappan1981thermocapillary} for the limiting case of a creeping flow and a small capillary, as well as Marangoni numbers. 
If the capillary number increases, the drop undergoes a deformation leading to a different interaction with the wall. Ascoli and Leal \cite{ascoli1990thermocapillary} studied such a scenario using the Boundary Integral Method (BIM). They also focused on the creeping flow regime with large thermal diffusivities, that is, small Reynolds and Marangoni numbers.
We devise a similar test case to study interactions of a Marangoni flow driven droplet with a plane wall, for which the numerical setup is shown in figure \ref{fig:deformable}$a$. For this case, the choice of dimensionless numbers is the same as in Section \ref{sec:ygb} with an additional parameter, i.e. the dimensionless distance of the drop centroid from the wall $D/R$. We fix the Reynolds number $\textrm{Re} = 0.08$, the Marangoni number $\textrm{Ma} = 10^{-6}$ and investigate the effect of two capillary numbers $\textrm{Ca}$ = $0.15$ and $\textrm{Ca} = 4.5$. It is assumed that $\mu_1 = 100 \mu_2$ and $\alpha_1 = \alpha_2$.

In figure \ref{fig:deformable}$b$, we show the velocity of the drop centroid $u$ non-dimensionalized with equation \eqref{eq:ygb} as a function of the distance of its centroid from the wall for $\textrm{Ca} = 0.15$. In this case, the drop remains spherical as it approaches the wall and our numerical results from all three methods agree well with the analytical calculations of Meyappan \textit{et al.} \cite{meyyappan1981thermocapillary}. Similarly to the previous cases, the height function method suffers from spurious oscillations because of the switch between vertical/horizontal height functions (see the drop migration movies at \url{http://basilisk.fr/sandbox/msaini/Marangoni/marangoniwall.c}).

\begin{figure}[h!]
    \centering
    \includegraphics[scale = 0.9]{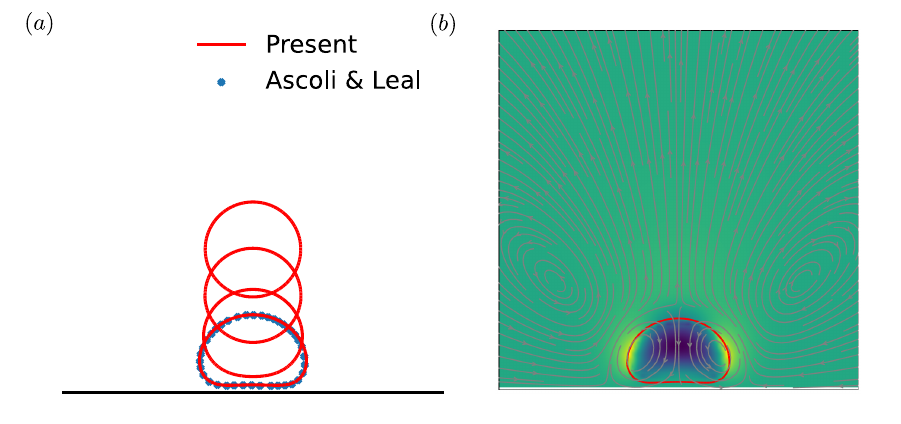}
    \caption{Results obtained from the CLSVOF method. $(a)$ The evolution of drop shape as it approaches the plane wall for $\textrm{Ca} = 4.5$, {\color{red} Marangoni number $\rm{Ma} = 10^{-6}$, Reynolds number $\rm{Re} = 0.066$} and the comparison with Ascoli and Leal \cite{ascoli1990thermocapillary} at dimensionless time $t \gamma_T \nabla T /\mu R \in \{2,4,6,8\}$. $(b)$ Velocity magnitude and streamlines near the drop.}
    \label{fig:defshape}
\end{figure}

In figure \ref{fig:deformable}$c$, the dimensionless migration velocity of the drop is shown for the case of $\textrm{Ca} = 4.5$. The present numerical results agree well with those of Ascoli and Leal \cite{ascoli1990thermocapillary}. Compared to the previous case of $\textrm{Ca} = 0.15$, the numerical oscillations for the HF method are smaller because the surface tension force is relatively low. The evolution of the drop shape as well as a one-to-one comparison with Ascoli and Leal's \cite{ascoli1990thermocapillary} final drop shape are shown in figure \ref{fig:defshape}$a$. The Marangoni flows around the droplet are depicted in Fig. \ref{fig:defshape}$b$. In the latter figure, we only show the results predicted by the CLSVOF method, as for all methods the flow fields are identical.

\subsection{Interaction of two unequal drops translating due to a Marangoni flow}

\begin{figure}
    \centering
    \includegraphics[scale = 0.8]{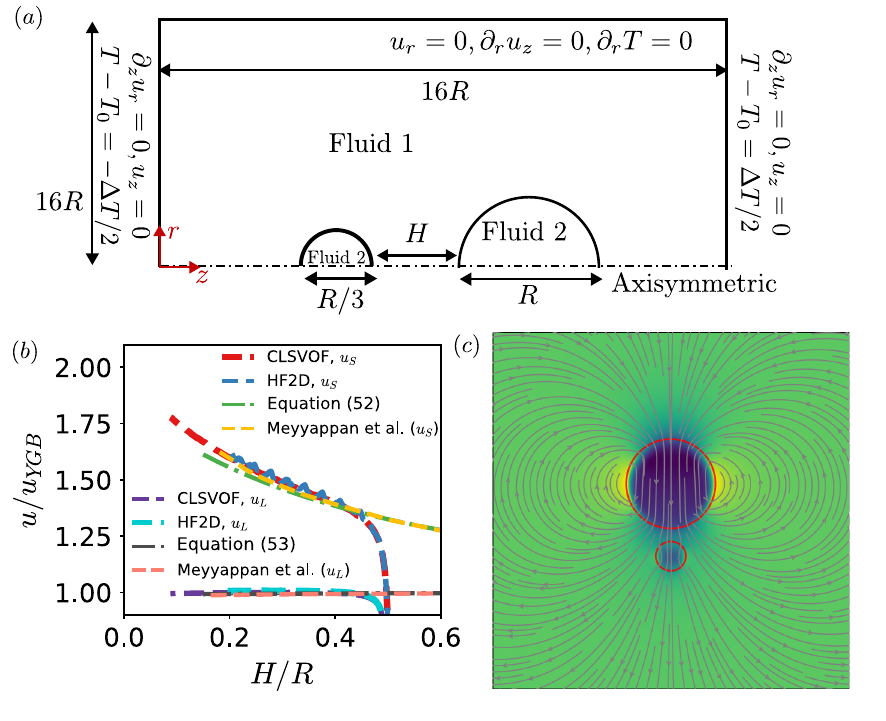}
    \caption{Interaction of two Marangoni drops for {\color{red} Marangoni number $\rm{Ma} = 10^{-6}$, Reynolds number $\rm{Re} = 0.066$ and Capillary number $\rm{Ca} = 0.25$} $(a)$ Numerical setup $(b)$ The velocity of both large and small drops $u_L$ and $u_S$ respectively for the CLSVOF and the HF2D methods. Comparison of the present results with equations \eqref{eq:vellarge} and \eqref{eq:velsmall} as well as with the accurate numerical solution of Meyyappan \textit{et al.} \cite{meyyappan1983slow}. $(c)$ The flow field and the streamlines near the Marangoni drops as obtained from the CLSVOF method. The code to reproduce the results in this figure is available at \url{http://basilisk.fr/sandbox/msaini/Marangoni/marangoniad2bub.c}. }
    \label{fig:2bubfield}
\end{figure}

The interaction of two droplets of unequal size driven by a Marangoni flow was studied analytically by Meyyappan \textit{et al.} \cite{meyyappan1983slow}. The herein developed solvers are used to study this interaction for the configuration shown in figure \ref{fig:2bubfield}$a$ where the two drops of different radii ($R_1$ and $R_2$) are placed in a temperature field that obeys equation \eqref{eq:advdiffT}. The domain size, boundary conditions, and choice of dimensionless parameters are the same as for section \eqref{sec:ygb} with two additional parameters: the size ratio of the two drops $R_1/R_2$ and the dimensionless gap between them $H/R$. The Capillary number is set to $\textrm{Ca} = 0.25$, the Reynolds number to $\textrm{Re} = 0.066$ and the Marangoni number to $\textrm{Ma} = 10^{-6}$. The viscosity and thermal diffusivity of the bulk fluid is much higher than those of the droplets, that is, $\mu_1 = 2000 \mu_2$ and $\alpha_1 = 2000 \alpha_2$. We investigate a particular case in which $R_1 = R$ and $R_2 = R/3$. The gap between the two drops is initially $R/2$. The surface tension decreases as the temperature increases; therefore, a Marangoni flow is generated, causing the droplets to migrate towards the cold boundary (see fig. \ref{fig:2bubfield} $a$). 
The larger droplet induces a higher migration velocity of the smaller drop $u_S$ such that the increase compared to Young's prediction (equation \eqref{eq:ygb}) is approximately


\begin{equation}
    \frac{u_S}{u_{YGB}} = 1 + 2 \left(\frac{3R}{3H + 4R}\right)^3,
    \label{eq:velsmall}
\end{equation}

\noindent whereas the velocity of the larger droplet $u_L$ decreases slightly to approximately


\begin{equation}
    \frac{u_L}{u_{YGB}} = 1 - \frac{2}{3} \left(\frac{R}{3H + 4R}\right)^3.
    \label{eq:vellarge}
\end{equation}

\noindent Equations \eqref{eq:velsmall} and \eqref{eq:vellarge} are obtained in the limit of when the size of the leading drop is much smaller compared to the trailing one ($R_2 \ll R_1$). Meyyappan \textit{et al.} \cite{meyyappan1983slow} also obtained accurate numerical results for the case of a finite-size leading droplet. In figure \ref{fig:2bubfield}$b$ we compare the present results of HF2D and CLSVOF with the numerical simulations of Meyyappan \textit{et al.} \cite{meyyappan1983slow}, and with equations \eqref{eq:velsmall} and \eqref{eq:vellarge}. After the initial transient phase, the results agree well with the numerical results of Meyappan \textit{et al.} \cite{meyyappan1983slow} while there are some differences compared to equations \eqref{eq:velsmall} and \eqref{eq:vellarge} due to the finite size of the smaller droplet. Similar differences are reported by Meyyappan \textit{et al.}. The results from the HF method have very large spurious oscillations and blow up for this case, so they are not included. Finally, the velocity field and streamlines around the two drops are also shown in figure \ref{fig:2bubfield}$c$.



\section{Conclusions and future outlook \label{sec:conclusions}}
The numerical methods developed in this work represent a significant advancement in the simulation of flows driven by surface tension, especially for applications involving variable surface tension. The coupled level set volume of
fluid (CLSVOF), height function (HF), and height function to distance (HF2D) approaches each bring distinct strengths, offering us a versatile set of tools for studying complex multiphase flow problems such as advection induced symmetry breaking, evaporation of droplets etc. 
These methods rely on a discretization of the surface tension force that conserves momentum, both globally and locally. In particular, the CLSVOF method is shown to be the most accurate and stable method with the lowest numerical oscillations compared to HF and HF2D. This property is critical for accurate numerical modeling of Marangoni flows. However, CLSVOF involves more numerical operations as compared to HF and HF2D, and is therefore relatively slower. {\color{blue}We also note that, to our knowledge, these methods are the first which are both mass and momentum conserving, an important property when describing surface-tension-driven two-phase flows, as discussed in \cite{popinet2018numerical}.}

The current implementations are limited to 2D or axisymmetric flows; thus, an important future outlook is to extend these implementations to account for the third spatial dimension. In the future, we also plan to extensively use recently developed numerical methods to investigate the problems of Marangoni flows, for example autothermotaxis, evaporating binary droplets, and many more.

\section*{Appendix}
\label{sec:sample:appendix}
\noindent \emph{(A) Computational time elapsed in different parts of the solvers} \\
Table \ref{Tab:timepercent} shows the percentage of total time spent in various parts of the solvers. For all the methods, more than 50 \% of the time is spent in dynamic regridding and interpolation of various quantities from coarse to refined levels or vice versa. The time spent solving the Poisson equation is nearly 21-27 \%, depending on the solver. For the calculation of the surface tension force using the CLSVOF method, around 18.2 \% of the time is used. Within that, around 93 \% of the time elapses during the redistancing of the level set function; however, only 7 \% is spent on other operations related to the calculation of the surface tension. The time required for the surface tension calculation is significantly lower for the HF method (8.8 \%) and the HF2D method (12.1 \%) as there is no requirement for the redistancing of the level set. Approximately 4 \% of the time is spent in operations related to the VOF method, that is, advection and reconstruction. Finally, around 6\% of the time is spent in the rest of the operations such as averaging various quantities at the interface, calculation of the viscous effects, etc.

\begin{table}[]
\resizebox{\columnwidth}{!}{%
\begin{tabular}{|l|l|l|l|}
\hline
                                 & CLSVOF  & HF      & HF2D    \\ \hline
Regridding and interpolations    & 50.9 \% & 55.5 \% & 52.6 \% \\ \hline
Pressure Poisson equation        & 21.1 \% & 26.6 \% & 24.9 \% \\ \hline
Surface tension calculation      & 18.2 \% & 8.8 \%  & 12.1 \% \\ \hline
VOF advection and reconstruction & 3.9 \%  & 3.7 \%  & 4.5 \%  \\ \hline
Other operations                 & 5.9 \%  & 5.4 \%  & 5.9 \%  \\ \hline
\end{tabular}%
}
\caption{Percentage of time spent in various parts of the solver.}
\label{Tab:timepercent}
\end{table}

\vspace{0.5cm}
\noindent {\bf Authors' contribution}: M.S. - Conceptualization, coding, validation and testing, discussion, writing original draft, review, and editing. V.S. - discussion, review, and editing. Y.S. - discussion, review, and editing. D.L. - Conceptualization, discussion, supervision, funding acquisition, review, and editing. S.P. - Conceptualization, coding, validation, discussion, review, and editing.

\noindent {\bf Acknowledgments}: This work was supported by an Industrial Partnership Programme of the Netherlands Organisation for Scientific Research (NWO) \& High Tech Systems and Materials (HTSM), co-financed by Canon Production Printing Netherlands B.V., University of Twente, and Eindhoven University of Technology. 

 \bibliographystyle{elsarticle-num} 
 \bibliography{cas-refs}





\end{document}